\documentclass[12pt]{article}
\usepackage{amssymb}
\usepackage{amsmath}
\usepackage{xypic}
\usepackage{graphicx}
\usepackage{pdfsync}

\newtheorem{theorem}{Theorem}

\newtheorem{prop}[theorem]{Proposition}

\newtheorem{lemma}[theorem]{Lemma}

\newtheorem{cor}[theorem]{Corollary}

\newtheorem{exam}[theorem]{Example}

\newtheorem{remark}[theorem]{Remark}

\newcommand{\eqa}{\begin{eqnarray}}
\newcommand{\eeqa}{\end{eqnarray}}
\newcommand{\beq}{\begin{equation}}
\newcommand{\eeq}{\end{equation}}
\newcommand{\nn}{\nonumber}

\newcommand{\pal}{\partial}

\newcommand{\pf}{\noindent{\it Proof \ }}
\newcommand{\tr}{{\rm tr}}

\newcommand{\epf}{$\quad$\hfill
\raisebox{0.11truecm}{\fbox{}}\par\vskip0.4truecm}

\title{Gromov--Witten invariants and integrable hierarchies of topological type}
\author{{B.Dubrovin$^*$}\\
{\small $^*$ SISSA, Trieste,
and 
Steklov Mathematical Institute, Moscow }}
\date{}
\begin{document}
\maketitle

\begin{flushright}
\parbox{9cm}{
\begin{center}
{\it Dedicated
to Sergei Petrovich Novikov \\
on the occasion of his birthday.}
\end{center}
}
\end{flushright}

\begin{abstract} We outline two approaches to the construction of integrable hierarchies associated with the theory of Gromov--Witten invariants of smooth projective varieties. We argue that a comparison of these two approaches yields nontrivial constraints on Chern numbers of varieties with semisimple quantum cohomology.
\end{abstract}

%\tableofcontents

\setcounter{equation}{0}
\setcounter{theorem}{0}
\section{Introduction}\par

The discovery of deep connections between topology of the Deligne--Mumford moduli spaces of stable algebraic curves and Korteweg--de Vries (KdV) hierarchy \cite{Witten2}, \cite{konts}, \cite{okun}, \cite{mirza}, \cite{kl} revealed a new unexpected face of the theory of integrable systems. A program of describing the intersection theory on more general moduli spaces of stable maps of algebraic curves to smooth projective varieties in terms of suitable integrable hierarchies first outlined in \cite{Witten2} remains one of the main challenges on the way towards better understanding of relationships between the theory of integrable systems and topology. It was however soon realized that, except for few low-dimensional cases (see more details at the end of Section \ref{sect33}) the needed integrable hierarchies never appeared in the theory of integrable systems. 

One arrives at the following question: given a smooth projective variety (or, more generally speaking, a compact symplectic manifold) $X$, can one construct a hierarchy of integrable partial differential equations (PDEs) such that the generating function of intersection numbers on the moduli stacks of stable maps of punctured algebraic curves to $X$ is given by (logarithm of) the tau-function of a particular solution to the hierarchy? An affirmative answer to this question was obtained in \cite{DZ} under the assumption of semisimplicity of quantum cohomology of $X$. The basic idea of the construction that already appeared in \cite{dz98} is very simple. Namely, we start with the genus zero hierarchy constructed in \cite{D92} (see also \cite{wit}) and, then, apply the so-called \emph{quasitriviality} substitution changing the dependent variables of the hierarchy. The main question is how to find such a substitution. The original approach suggested in \cite{dz98} uses the topological approach to the construction of the total descendent Gromov--Witten (GW) potential based on the study of identities \cite{getzler1} in the tautological ring of the Deligne--Mumford spaces (cf. \cite{giv98}). An alternative approach of \cite{DZ} uses expressions for the genus $g\geq 1$ total GW potential via the genus zero quantities derived from the \emph{Virasoro constraints}, following the idea of  \cite{ehx}. In order to prove uniqueness of a solution to the system of Virasoro constraints the semisimplicity of quantum cohomology of $X$ was used in \cite{DZ}

The main goal of this paper is to compare these two approaches. We show that the result of such a comparison are not completely trivial already in the study of the degree zero GW invariants, as it implies a constraint on the Chern numbers of a smooth projective variety with semisimple quantum cohomology (see Proposition \ref{prop45} below and also \cite{libgober}, \cite{ehx}, \cite{borisov}\footnote{The author thanks Burt Totaro and Hsian-Hua Tseng for bringing his attention to the papers \cite{libgober} and \cite{borisov}.}).

The paper is organized as follows. In Section \ref{sect2} we outline the basics of our construction of integrable hierarchies applying it to the well known example of the KdV hierarchy. Generalizing this construction we introduce in Section \ref{sect3} an \emph{integrable hierarchy of topological type} associated with an arbitrary smooth projective variety $X$ with $H^{\rm odd}(X, \mathbb C)=0$ having semisimple quantum cohomology, adapting to the topological setting the main ideas of \cite{DZ}. The last Section \ref{sect4} is dedicated to the comparison of results of the Section \ref{sect3} with simple properties of moduli stacks of stable maps of degree zero.

\section{A construction of KdV hierarchy}\label{sect2}

\subsection{KdV hierarchy from Lax representation}.

It is well known that the Korteweg--de Vries (KdV) equation for a function $u=u(x,t)$
\beq\label{kdv1}
u_t =u\, u_x +\frac{\epsilon^2}{12} u_{xxx}
\eeq
(the subscripts stand for partial derivatives) makes part of an infinite family of evolutionary partial differential equations (PDEs) for a function $u=u(x; t_0, t_1, t_2, \dots)$, the so-called \emph{KdV hierarchy}. The first few equations of the hierarchy read
\eqa\label{hier1}
&&
u_{t_0}=u_x
\nn\\
&&
u_{t_1} =u\, u_x +\frac{\epsilon^2}{12} u_{xxx}
\\
&&
u_{t_2}=\frac{u^2}{2!} u_x +\frac{\epsilon^2}{12} \left( 2 u_x u_{xx}+u \, u_{xxx}\right)+\frac{\epsilon^4}{240}u^{V}
\nn
\eeqa
etc. In a nutshell \emph{integrability} of equations of the hierarchy means validity of the compatibility identities
\beq\label{compat}
\left( u_{t_i}\right)_{t_j}=\left( u_{t_j}\right)_{t_i}
\eeq
for every pair of indices $i$, $j$ plus certain condition of completeness of the family of commuting flows \eqref{hier1} that we will not discuss here. The small parameter $\epsilon$ is responsible for dispersive effects of solutions to the KdV equation. Although it can be eliminated by rescaling $x\to \epsilon\, x$, $t_k \to \epsilon\, t_k$, we prefer to keep it as it will reappear below in the genus expansion of solutions to hierarchies of topological type.

There are various constructions of the KdV hierarchy. The most well known uses fractional powers of the Lax operator
\beq\label{lax}
L=\frac12 (\epsilon\, \pal_x)^2 +u(x).
\eeq
Define differential operators $A_i$ by the formula
\beq\label{lax1}
A_i = \frac1{(2i+1)!!} \left(2 L\right)^{\frac{2i+1}2}_+=
\frac1{(2i+1)!!} \left[ (\epsilon\,\pal_x)^{2i+1} +(2i+1) u(x)(\epsilon\, \pal_x)^{2i-1}+\dots\right].
\eeq
Here $(~)_+$ refers to the differential part of the pseudodifferential operator $\left(2 L\right)^{\frac{2i+1}2}$. The $i$-th equation of the hierarchy admits the Lax representation
\beq\label{lax2}
u_{t_i}\equiv L_{t_i} = [A_i, L].
\eeq
%The KdV itself corresponds to $i=1$.

\subsection{KdV hierarchy by a quasitriviality transformation}\par

Let us briefly outline one more construction of the KdV hierarchy, perhaps the one less known than others. Start with the so-called \emph{dispersionless limit} of the equations \eqref{hier1}
\eqa\label{hier0}
&&
v_{t_0}=v_x
\nn\\
&&
v_{t_1} =v\, v_x
\\
&&
v_{t_2}=\frac{v^2}{2!} v_x
\nn\\
&&
\dots \dots \dots
\nn\\
&&
v_{t_i} =\frac{v^i}{i!} v_x
\nn
\eeqa
obtained by setting $\epsilon$ to zero (we have redenoted $u\to v$ the dependent function of the hierarchy for a later convenience). This is also an integrable hierarchy: validity of the compatibility identities
\beq\label{compat0}
\left( v_{t_i}\right)_{t_j} = \left( v_{t_j}\right)_{t_i}
\eeq
can be checked by a one line computation. We want to reconstruct the full hierarchy \eqref{hier1} from its dispersionless limit \eqref{hier0}. To this end perform a substitution (the so-called \emph{quasitriviality transformation})
\beq\label{quasi1}
v\mapsto u =v+{\epsilon^2\over 24} \left( \log v_x\right)_{xx}
+\epsilon^4 \left( {v_{xxxx}\over 1152\, {v_x}^2}
- {7\, v_{xx} v_{xxx}\over 1920\, {v_x}^3}
+{{v_{xx}}^3\over 360\, {v_x}^4}\right)_{xx} + O(\epsilon^6)
\eeq
Let us explain the procedure at the first order in $\epsilon^2$. One has
\eqa\label{kdv4}
&&
u_{t_i} = v_{t_i} +\frac{\epsilon^2}{24} \left(\frac{v_{x\, t_i}}{v_x}\right)_{xx} +{\mathcal O}\left(\epsilon^4\right)= \frac{\partial}{\partial x}\left[\frac{v^{i+1}}{(i+1)!} +\frac{\epsilon^2}{24} \left( \frac{v^i}{i!} \frac{v_{xx}}{v_x} + \frac{v^{i-1}}{(i-1)! }v_x\right)_x\right]+\dots
\nn\\
&&
=\frac{\pal}{\pal x}\left[ \frac{u^{i+1}}{(i+1)!} +\frac{\epsilon^2}{24} \left( 2\frac{u^{i-1}}{(i-1)!} u_{xx} +\frac{u^{i-2}}{(i-2)!} u_x^2\right)\right] +{\mathcal O}\left(\epsilon^4\right).
\eeqa
It can be easily checked that the resulting equation coincides with \eqref{lax2}, within the $\epsilon^2$ approximation.

Integrability \eqref{compat} of the full KdV hierarchy now readily follows from integrability \eqref{compat0} of the very simple hierarchy \eqref{hier0}. Indeed, a change of variables preserves commutativity of the flows.

One can apply the same substitution to the bihamiltonian structure of the hierarchy \eqref{hier0}. The latter is given by a compatible pair of Poisson brackets
\eqa\label{pb1}
&&
\{ v(x), v(y)\}_1= \delta'(x-y)
\\
&&
\{ v(x), v(y)\}_2=v(x) \delta'(x-y) +\frac12 v_x\delta(x-y).
\label{pb2}
\eeqa
Recall that compatibility of the Poisson brackets means that the linear combination
\beq\label{plam}
\{ v(x), v(y)\}_\lambda:= \{ v(x)-v(y)\}_2 -\lambda\{ v(x), v(y)\}_1
\eeq
is again a Poisson bracket for any $\lambda$. Equations of the hierarchy \eqref{hier0} can be represented in the Hamiltonian form in two different ways
\eqa\label{biham0}
&&
v_{t_i}=\{ v(x), H_i\}_1 =\pal_x \frac{\delta H_i}{\delta v(x)}
\\
&&
\quad\quad
=\left( i+\frac12\right)^{-1} \{ v(x), H_{i-1}\}_2 =\left( v\pal_x +\frac12 v_x\right)\frac{\delta H_{i-1}}{\delta v(x)}
\nn
\eeqa
where the Hamiltonian is defined by
\beq\label{ham0}
H_i =\int \frac{[v(x)]^{i+2}}{(i+2)!}\, dx.
\eeq
Applying the same substitution to the Poisson brackets \eqref{pb1}, \eqref{pb2} and to the Hamiltonians \eqref{ham0} we arrive at a bihamiltonian structure of the full KdV hierarchy. Again, let us explain the idea of the calculation computing the first correction in $\epsilon^2$. Begin with the Hamiltonian. Within the ${\mathcal O}\left( \epsilon^2\right)$ order one has
\beq\label{suv}
v=u-\frac{\epsilon^2}{24} \left( \log u_x\right)_{xx}+{\mathcal O}\left( \epsilon^4\right).
\eeq
So the Hamiltonian density
$\frac{v^{i+2}}{(i+2)!}$ after the substitution becomes
\eqa
&&
\frac{v^{i+2}}{(i+2)!}=\frac{u^{i+2}}{(i+2)!} -\frac{\epsilon^2}{24} \frac{u^{i+1}}{(i+1)!}  \left( \log u_x\right)_{xx}+{\mathcal O}\left( \epsilon^4\right)
\nn\\
&&
=\frac{u^{i+2}}{(i+2)!}-\frac{\epsilon^2}{24}\left[\frac{u^{i-1}}{(i-1)!} u_x^2+\pal_x\left( \frac{u^{i+1}}{(i+1)!}\frac{u_{xx}}{u_x} - \frac{u^i}{i!} u_x\right)\right]+{\mathcal O}\left( \epsilon^4\right)
\nn
\eeqa
The total $x$-derivative does not contribute to the Hamiltonian. Therefore 
$$
H_i =\int \left[ \frac{u^{i+2}}{(i+2)!}-\frac{\epsilon^2}{24}\frac{u^{i-1}}{(i-1)!} u_x^2\right] \, dx+{\mathcal O}\left( \epsilon^4\right).
$$

Actually there is a more smart way for calculating the Hamiltonian densities that already eliminates all non-polynomial expressions. Besides the substitution \eqref{suv} add a total derivative
\beq\label{dense}
h_i:=\frac{v^{i+2}}{(i+2)!} +\epsilon^2 \frac{\pal^2 \Delta{\cal F}}{\pal x\,\pal t_{i+1}}=\frac1{(i+2)!} \left[ u-\frac{\epsilon^2}{24} \left( \log u_x\right)_{xx}\right]^{i+2} +\frac{\epsilon^2}{24} \frac{\pal^2 \log u_x}{\pal x \,\pal t_{i+1}}+{\mathcal O}\left( \epsilon^4\right).
\eeq
Here
$$
\Delta{\cal F}={1\over 24} \log v_x
+\epsilon^2 \left( {v_{xxxx}\over 1152\, {v_x}^2}
- {7\, v_{xx} v_{xxx}\over 1920\, {v_x}^3}
+{{v_{xx}}^3\over 360\, {v_x}^4}\right) + O(\epsilon^4)
$$
(cf. \eqref{quasi1}). After simple calculation one obtains a polynomial, at this order, Hamiltonian density
$$
h_i=\frac{u^{i+2}}{(i+2)!} +\frac{\epsilon^2}{24}\left(2 \frac{u^i u_{xx}}{i!} +\frac{u^{i-1} u_x^2}{(i-1)!} \right) +{\mathcal O}\left( \epsilon^4\right).
$$
The Hamiltonian
$
H_i =\int h_i\, dx
$
coincides with the one obtained above but the densities differ by a total $x$-derivative. An advantage of the Hamiltonian densities $h_i=h_i(u, u_x, \dots; \epsilon)$ defined by \eqref{dense}\footnote{A construction of the Hamiltonian densities of the KdV hierarchy satisfying the tau-symmetry condition \eqref{taus1} in terms of the Hadamard--Seeley coefficients of the Lax operator \eqref{lax} can be found in \cite{DZ}.} is in validity of the following identities
\beq\label{taus1}
\frac{\pal h_{i-1}}{\pal t_j} = \frac{\pal h_{j-1}}{\pal t_i}
\eeq
for arbitrary $i$, $j\geq 0$. Due to the \emph{tau-symmetry} \eqref{taus1} one concludes that, for any common solution $u=u(x, t_0, t_1, \dots)$ of the KdV hierarchy there exists a function $\tau=\tau(x, t_0, t_1, \dots)$ such that
\beq\label{taus2}
h_i(u, u_x, \dots) =\epsilon^2 \frac{\pal^2 \log \tau}{\pal x\, \pal t_{i+1}}.
\eeq
Such a representation makes sense also for $i=-1$ giving the trivial Hamiltonian
$$
h_{-1}= u =\epsilon^2 \frac{\pal^2\log\tau}{\pal x^2}
$$
(the Casimir of the first Hamiltonian structure of the KdV hierarchy (see below)).

The transformation of the Poisson brackets can be calculated in a straightforward way. Let us apply the substitution \eqref{quasi1} directly to the pencil \eqref{plam} of Poisson brackets. One has
\eqa
&&
\{ u(x), u(y)\}_\lambda=\{ v(x)+\frac{\epsilon^2}{24} (\log v_x)_{xx}, v(y)+\frac{\epsilon^2}{24} (\log v_y)_{yy}\}_\lambda+\dots
\nn\\
&&
\{ v(x), v(y)\}_\lambda +\frac{\epsilon^2}{24} \left[ \{ v(x), (\log v_y)_{yy}\}_\lambda +\{ (\log v_x)_{xx}, v(y)\}_\lambda\right]+\dots
\nn\\
&&
=\{ v(x), v(y)\}_\lambda +\frac{\epsilon^2}{24} \left[ \pal_y^2 v_y^{-1} \pal_y \{ v(x), v(y)\}_\lambda +\pal_x^2 v_x^{-1} \pal_x \{ v(x), v(y)\}_\lambda \right]+\dots .
\label{skob}
\eeqa
Using explicit expression for
$$
\{ v(x), v(y)\}_\lambda=\left( v(x)-\lambda\right) \delta'(x-y) +\frac12 v_x \delta(x-y)
$$
along with the general identity for derivatives of delta-function
\beq\label{delta}
f(y) \delta^{(k)}(x-y)=f(x) \delta^{(k)}(x-y)+\sum_{j=1}^k \left( \begin{array}{c} k \\ j\end{array}\right) f^{(j)}(x) \delta^{(k-j)}(x-y)
\eeq
one rewrites the expression \eqref{skob} in the form
\eqa
&&
\{ u(x), u(y)\}_\lambda=\left( v(x)-\lambda\right) \delta'(x-y) +\frac12 v_x \delta(x-y) + \frac{\epsilon^2}{24}\left[ (\log v_x)_{xx} \delta'(x-y) 
\right.
\nn\\
&&\left.
+\frac12 (\log v_x)_{xxx} \delta(x-y)+3\delta'''(x-y)\right]+\dots 
\nn\\
&&
=\left( u(x)-\lambda\right) \delta'(x-y) +\frac12 u_x \delta(x-y) +\frac{\epsilon^2}8 \delta'''(x-y)+{\mathcal O}\left( \epsilon^4\right).
\nn
\eeqa
It can be shown that terms of higher order in this expression all vanish. 

\begin{remark} The substitution \eqref{quasi1} transforms solutions to equations of the dispersionless hierarchy \eqref{hier0} to those of the full KdV hierarchy \eqref{hier1}. The resulting solution $u=u(x; {\bf t}; \epsilon)$ will be written as a series in $\epsilon^2$. However, it is applicable only to solutions of \eqref{hier0} monotone in $x$ due to presence of $v_x$ in the denominators\footnote{It turns out that higher jets $v^{(k)}$ for $k>1$ do not appear in the denominators.}.
\end{remark}

\subsection{Loop equation in the KdV theory}

Let us now outline the construction of the substitution \eqref{quasi1}. Define a sequence of second order linear differential operators ${\mathcal L}_m$, $m\geq -1$ with coefficients depending on a parameter $\epsilon$ acting on functions of jet variables $v\equiv v^{(0)}$, $v_x\equiv v^{(1)}$, $v_{xx}\equiv v^{(2)}$, \dots, $v^{(k)}$, \dots . They are conveniently defined by the following generating function
\beq\label{jvir1}
\sum_{m=-1}^\infty \frac{{\mathcal L}_m}{\lambda^{m+2}}=
\sum_{k\geq 0} \left(A_k(\lambda) -\epsilon^2 B_k(\lambda)\right)\frac{\pal}{\pal v^{(k)}} -\frac{\epsilon^2}2 \sum_{k,\, l\geq 0} C_{kl}(\lambda)\frac{\pal^2}{\pal v^{(k)} \pal v^{(l)}}+\frac1{16(v-\lambda)^2}
\eeq
where
\eqa\label{jvir2}
&&
A_k(\lambda)=A_k\left(\lambda; v, v_x, \dots, v^{(k)}\right) =\pal_x^k \left(\frac1{v-\lambda}\right)+\sum_{j=1}^k \left(\begin{array}{c} k \\ j\end{array}\right)\pal_x^{j-1}\left(\frac1{\sqrt{v-\lambda}}\right) \pal_x^{k-j+1}\left(\frac1{\sqrt{v-\lambda}}\right)
\nn\\
&&
B_k(\lambda)=B_k\left(\lambda; v, v_x, \dots, v^{(k+2)}\right) =-\frac1{16} \pal_x^{k+2} \left( \frac1{(v-\lambda)^2}\right)
\\
&&
C_{kl}(\lambda)=C_{kl}\left(\lambda; v, v_x, \dots, v^{(\max (k, l)+1)}\right) = \pal_x^{k+1} \left(\frac1{\sqrt{v-\lambda}}\right) \pal_x^{l+1} \left(\frac1{\sqrt{v-\lambda}}\right)
\nn
\eeqa
and the operator $\pal_x$ is defined by
\beq\label{palx}
\pal_x=\sum_{k\geq 0} v^{(k+1)} \frac{\pal}{\pal v^{(k)}}.
\eeq
Clearly the coefficients $A_k(\lambda)$, $B_k(\lambda)$, $C_{kl}(\lambda)$ are rational functions of $\lambda$ with poles only at $\lambda=v$. They depend polynomially on the jets $v_x$, $v_{xx}$ etc.  Explicitly,
\eqa
&&
{\mathcal L}_{-1}=-\frac{\pal}{\pal v}
\nn\\
&&
{\mathcal L}_0=-\sum_{k\geq 0} \frac{2k+1}2 v^{(k)} \frac{\pal}{\pal v^{(k)}}+\frac1{16}
\nn
\eeqa
etc.

\begin{lemma} The operators \eqref{jvir1} satisfy commutation relations
\beq\label{jvir0}
[{\mathcal L}_i, {\mathcal L}_j] = (i-j) {\mathcal L}_{i+j}, \quad i, \, j\geq -1.
\eeq
\end{lemma}

We are now ready to write a system of equations defining the substitution \eqref{quasi1}. 

\begin{theorem} There exists a unique, up to additive constants,  sequence of functions ${\cal F}_g\left(v, v_x, \dots, v^{(3g-2)}\right)$ satisfying the following system of equations
\beq\label{vircons}
{\mathcal L}_m e^{\Delta {\cal F}}=0, \quad m\geq -1
\eeq
where
\beq\label{deltaf}
\Delta{\cal F}:= \sum_{g\geq -1} \epsilon^{2g-2} {\cal F}_g\left(v, v_x, \dots, v^{(3g-2)}\right).
\eeq
\end{theorem}

The equations \eqref{vircons} yield
\beq\label{fen1}
{\cal F}_1=\frac1{24} \log v_x
\eeq
and then an overdetermined system of recursion relations
\beq\label{virec}
\sum_{k=0}^{3g-2} A_k(\lambda) \frac{\pal {\cal F}_g}{\pal v^{(k)}}=\sum_{k=0}^{3g-5} B_k(\lambda)  \frac{\pal {\cal F}_{g-1}}{\pal v^{(k)}}+\frac12\sum_{k, \, l=0}^{3g-5} C_{kl}(\lambda) \left( \frac{\pal^2 {\cal F}_{g-1}}{\pal v^{(k)} \pal v^{(l)}} +\sum_{h=1}^{g-1} \frac{\pal {\cal F}_h}{\pal v^{(k)}} \frac{\pal {\cal F}_{g-h}}{\pal v^{(l)}}\right)
\eeq
depending on the parameter $\lambda$. Compatibility of the resulting overdetermined linear system essentially follows from commutation relations \eqref{jvir0}. So one obtains
$$
{\cal F}_2 ={v_{xxxx}\over 1152\, {v_x}^2}
- {7\, v_{xx} v_{xxx}\over 1920\, {v_x}^3}
+{{v_{xx}}^3\over 360\, {v_x}^4}
$$
etc.

The last step is the following

\begin{theorem} After the substitution
\beq\label{jquasi}
v\mapsto u=v+\epsilon^2 \pal_x^2 \Delta {\cal F}
\eeq
the dispersionless hierarchy \eqref{hier0} transforms to the KdV hierarchy \eqref{hier1}, \eqref{lax2}. Here $\Delta {\cal F}$ is the solution to the system \eqref{vircons}.
\end{theorem}

\setcounter{equation}{0}
\setcounter{theorem}{0}
\section{Integrable hierarchies of topological type}\label{sect3}\par

In this section the construction of integrable hierarchies of topological type will be adapted to the environment of Gromov--Witten (GW) invariants we will begin with.

\subsection{Gromov--Witten invariants of a smooth projective varieties, their descendents and total Gromov--Witten potential}\par

Let $X$ be a smooth projective variety of complex dimension $d$. For simplicity it will be assumed that
\beq\label{odd}
H^{\rm odd}(X,{\mathbb C})=0.
\eeq
Denote $n=\dim H^*(X,{\mathbb C})$. Choose a basis $\gamma_1=1$, $\gamma_2$, \dots, $\gamma_n$ of cohomologies. Assume
\beq\label{qa}
\gamma_\alpha\in H^{2q_\alpha}(X), \quad q_1=0, \quad q_n=d.
\eeq
The class $\gamma_n\in H^{2d}(X)$ coincides with the volume element normalized by
\beq\label{vol}
\int_X\gamma_n=1.
\eeq
The Poincar\'e pairing on the cohomology will be denoted by
\beq\label{pair}
\langle a, b\rangle =\int_X a\wedge b.
\eeq
Denote
\beq\label{eta}
\eta_{\alpha\beta}=\int_X \gamma_\alpha\wedge \gamma_\beta
\eeq
the Gram matrix of the pairing with respect to the basis $\gamma_1$,  \dots, $\gamma_n$. We will use this symmetric matrix and its inverse $\eta^{\alpha\beta}$ for lowering and raising indices in the subsequent formulae.

GW invariants and their descendents can be defined as intersection numbers of certain cycles on the moduli stack $\overline{M}_{g,m} (X,\beta)$ of stable maps
\beq\label{mod1}
\overline{M}_{g,m} (X,\beta)=\{ f:\left( C_g; x_1, \dots, x_m\right) \to X,\quad f_* [C_g]=\beta\}/\mbox{equivalencies}
\eeq
of a given degree $\beta\in  H_2(X; \mathbb Z)/{\rm torsion}$.
Here $C_g$ is an algebraic curve of genus $g$ with at most double points singularities, $x_1$, \dots, $x_m$ are pairwise distinct marked points on $C_g$. Equivalencies are isomorphisms of $C_g \to X$ identical on $X$ and on the markings. Put
\beq\label{gw}
\left\langle \tau_{p_1}(\gamma_{\alpha_1})\dots \tau_{p_m}(\gamma_{\alpha_m})\right\rangle_{g,m,\beta}=\int_{\left[ \overline{M}_{g,m} (X,\beta)\right]^{\rm virt}}{\rm ev}_1^*(\gamma_{\alpha_1})\psi_1^{p_1}\dots {\rm ev}_m^*(\gamma_{\alpha_m})\psi_m^{p_m}.
\eeq
Here
\beq\label{ev}
{\rm ev}_i: \overline{M}_{g,m} (X,\beta)\to X, \quad f \mapsto f(x_i) 
\eeq
are evaluation maps,
\beq\label{psi}
\psi_i =c_1\left({\mathcal L}_i\right), \quad {\rm Chern ~ class ~ of ~tautological ~ line ~ bundle}\quad \begin{array}{c} {\mathcal L}_i  \\\Big\downarrow\vcenter{%
 \rlap{$\scriptstyle{T_{x_i}^*C_g}$}}   \\ \overline{M}_{g,m} (X,\beta)  \end{array}
\eeq
Finally, the \emph{virtual fundamental class} $\left[\overline{M}_{g,m} (X,\beta)\right] ^{\rm virt}$ is an element of the Chow ring $A_{*}\left(\overline{M}_{g,m} (X,\beta)\right)$
\beq\label{dimvirt}
\left[\overline{M}_{g,m} (X,\beta)\right] ^{\rm virt}\in A_D\left(\overline{M}_{g,m} (X,\beta)\right), \quad D=(1-g)(d-3) +m+\langle \beta, c_1(X)\rangle.
\eeq
Note that, due to \emph{effectivity} (see \cite{manin}) it suffices to consider the invariants \eqref{gw} only for cycles in the semigroup
\beq\label{eff}
\beta\in B(X)\subset H_2(X; \mathbb Z)/{\rm torsion}, \quad \int_\beta \omega \geq 0 \quad \forall \, \omega \in \, \mbox{ K\"ahler cone}\subset H^2(X).
\eeq

Generating function of rational numbers \eqref{gw}  (also called \emph{total descendent potential} of genus $g$) for a given genus $g$ is defined by the series
\beq\label{fg}
{\mathcal F}_g({\bf t}, q) =\sum_m  \sum_{(\alpha_1,p_1), \dots, (\alpha_m,p_m)} \frac{t^{\alpha_1}_{p_1}\dots t^{\alpha_m}_{p_m}}{m!}
\sum_{\beta\in H_2(X; \mathbb Z)} \left\langle \tau_{p_1}(\gamma_{\alpha_1})\dots \tau_{p_m}(\gamma_{\alpha_m})\right\rangle_{g,m,\beta} q^\beta.
\eeq
Here $t^\alpha_p$ are indeterminates, $\alpha=1, \dots, n$, $p=0$, $1$, \dots; 
\eqa
&&
q^\beta=q_1^{m_1}\dots q_l^{m_l} \quad\mbox{for}\quad  \beta=m_1\beta_1+\dots +m_l\beta_l
\nn\\
&& 
\nn\\
&& 
\mbox{where}\quad \beta_1, \dots \beta_l\quad \mbox{is a basis in}\quad H_2(X;\mathbb Z)/{\rm torsion}
\nn
\eeqa 
is an element of the Novikov ring. The total GW potential is defined by summation over genera,
\beq\label{total}
{\mathcal F}({\bf t},  q,\epsilon)=\sum_{g=0}^\infty \epsilon^{2g-2} {\mathcal F}_g.
\eeq
The additional parameter $\epsilon$ of the genus expansion is called \emph{string coupling constant} in physics literature. It will play the role of small dispersion expansion parameter in the integrable hierarchy we are going to construct under certain additional assumptions.

More specifically, our goal is to construct a hierarchy of integrable PDEs with $n$ dependent functions and time variables ${\bf t}=(t^\alpha_p)$, $t^1_0=x$, such that
$$
\tau({\bf t}, q, \epsilon)=e^{\mathcal F}
$$
is tau-function of a particular solution to the hierarchy.

In general it is hard to think about reconstruction of differential equations starting from just one solution. However, the design of construction explained in the previous section for the particular case of KdV hierarchy suggests a very natural recipe we are going to outline in this section.

The clue is in using the Frobenius structure on the manifold $M=H^*(X,{\mathbb C})$ defined by the genus zero $GW$ potential with no $\psi$-classes\footnote{No descendents, in physics terminology.}. Introduce affine coordinates $v^1$, \dots, $v^n$ on this space associated with the chosen basis $\gamma_1$, \dots, $\gamma_n$. Denote
$$
{\bf v}=v^\alpha \gamma_\alpha\in H^*(X,{\mathbb C})
$$
(summation over repeated indices here and below will be assumed). 
Put
\beq\label{pot}
F({\bf v},q)=\sum_{m} \sum_{\alpha_1, \dots, \alpha_m}\frac{v^{\alpha_1}\dots v^{\alpha_m}}{m!}\sum_{\beta\in H_2(X; \mathbb Z)} \int_{\overline{M}_{0,m}(X,\beta)} {\rm ev}_1^* (\gamma_{\alpha_1})\dots {\rm ev}_m^* (\gamma_{\alpha_m}) q^\beta.
\eeq
Due to \emph{divisor axiom} \cite{manin} the dependence on $q=(q_1, \dots, q_l)$ can be essentially eliminated by a shift
\eqa\label{shift}
&&
v^{\alpha_i}\mapsto v^{\alpha_i} - \sum_{s=1}^l M_{is} \log q_s, \quad i=1, \dots, l=\dim H^2(X), \quad M_{is}=\int_{\beta_s} \gamma_{\alpha_i}
\nn\\
&&
\gamma_{\alpha_1}, \dots, \gamma_{\alpha_l}\quad \mbox{a basis in}\quad H^2(X), \quad \beta_1, \dots, \beta_l \quad \mbox{a basis of the lattice}\quad H_2(X; \mathbb Z)/{\rm torsion}
\nn
\eeqa
under assumption of convergence. Such convergence will be assumed in sequel.
 So the dependence on $q=(q_1, \dots, q_l)$ is often omitted from the formulae of the theory of Frobenius manifolds.
Triple derivatives of this function
\beq\label{struc}
c_{\alpha\beta}^\gamma ({\bf v}) =\eta^{\gamma\, \delta}\frac{\partial^3 F({\bf v})}{\partial v^\delta \partial v^\alpha \partial v^\beta}
\eeq
where 
\beq\label{inv}
\left(\eta^{\alpha\beta}\right)=\left( \eta_{\alpha\beta}\right)^{-1},
\eeq 
define on $M$ a structure of Frobenius manifold, i.e., a structure of commutative associative algebra on the tangent space $T_{\bf v}M$ at every point ${\bf v}\in M$,
\beq\label{prod}
\frac{\pal}{\pal v^\alpha} \cdot \frac{\pal}{\pal v^\beta} =c_{\alpha\beta}^\gamma({\bf v}) \frac{\pal}{\pal v^\gamma},
\eeq
a flat invariant metric
\beq\label{metr}
\left\langle \frac{\pal}{\pal v^\alpha}, \frac{\pal}{\pal v^\beta}\right\rangle =\eta_{\alpha\beta}
\eeq
a unit
\beq\label{unit}
e=\frac{\pal}{\pal v^1}
\eeq
and an Euler vector field 
\beq\label{euler}
 E=\sum_{\alpha=1}^n \left[(1-q_\alpha) v^\alpha  + \left\langle c_1(X), \gamma^\alpha\right\rangle\right] \frac{\partial}{\partial v^\alpha}
 \eeq
entering into the \emph{quasihomogeneity} property of the potential $F$
\beq\label{quasi}
E\, F=(3-d) F+\mbox{quadratic polynomial}.
\eeq

\subsection{Genus zero invariants and Principal Hierarchy}\par

For any function $f({\bf t})$ of infinite number of variables $t^\alpha_p$ define an operation of \emph{restriction on the small phase space} by
\beq\label{small}
f({\bf t})|_{\rm small~ phase ~space}:= f({\bf t})|_{t^\alpha_0=v^\alpha, ~ t^\alpha_p=0~{\rm for}~ p>0, ~\alpha=1, \dots, n}.
\eeq
Thus, for example
$$
F({\bf v}, q)={\cal F}_0({\bf t}, q)|_{\rm small ~ phase ~ space}.
$$

Define functions $\theta_{\alpha,p}({\bf v},q)$ by
\beq\label{thetap}
\theta_{\alpha,p}({\bf v},q):= \frac{\pal ^2 {\cal F}_0({\bf t}, q)}{\pal t^1_0\, \pal t^\alpha_0}|_{\rm small~ phase ~space}.
\eeq
The generating functions
\beq\label{thetaz}
\theta_\alpha({\bf v}, q; z) =\sum_{p=0}^\infty \theta_{\alpha,p}({\bf v},q)z^p, \quad \alpha=1, \dots, n
\eeq
form a basis of horizontal sections
\beq\label{horizont}
\tilde\nabla(z) d\theta_\alpha({\bf v}, q; z)=0
\eeq
of the canonical \emph{deformed flat connection}
\beq\label{nabla}
\tilde\nabla_a(z)b =\nabla_ab +z\, a\cdot b, \quad a, \, b ~\mbox{are vector fields on }\quad M.
\eeq
A choice of such a basis is called \emph{calibration} of the Frobenius manifold $M$. In the GW setting we will always choose the calibration in the form \eqref{thetap}, \eqref{thetaz}.

We are ready to define \emph{Principal Hierarchy} associated with a calibrated Frobenius manifold as a system of PDEs for vector-valued functions ${\bf v}={\bf v}(x, {\bf t})$ taking values in $M=H^*(X)$. Equations of the hierarchy read
\beq\label{ph}
\frac{\partial {\bf v}}{\partial t^\alpha_p} =\frac{\partial}{\partial x} \nabla \theta_{\alpha, p+1}({\bf v},q), \quad \alpha=1, \dots, n, \quad p\geq 0.
\eeq 
In particular
\beq\label{t10}
\frac{\pal}{\pal t^1_0}=\frac{\pal}{\pal x}.
\eeq
The equation \eqref{ph} is a system of $n$ Hamiltonian PDEs with the Hamiltonian
\beq\label{hamph}
H_{\alpha, p}^0=\int \theta_{\alpha, p+1}({\bf v}(x)) \, dx
\eeq
and Poisson bracket
\beq\label{pbph}
\{ v^\alpha(x), v^\beta(y)\}_1=\eta^{\alpha\beta}\delta'(x-y).
\eeq
Moreover, the equations of the principal hierarchy commute pairwise, 
\beq\label{commute}
\{ H_{\alpha, p}^0, H_{\beta,q}^0\}_1=0\quad \Rightarrow\quad \frac{\pal}{\pal t^\alpha_p}\frac{\pal {\bf v}}{\pal t^\beta_q}=\frac{\pal }{\pal t^\beta_q}\frac{\pal {\bf v}}{\pal t^\alpha_p} =0.
\eeq
They also admit a second Hamiltonian structure defined by the Poisson bracket
\eqa\label{vtor}
&&
\{ v^\alpha(x), v^\beta(y)\}_2 =g^{\alpha\beta}({\bf v}(x)) \delta'(x-y) +\Gamma^{\alpha\beta}_\gamma({\bf v}(x)) \delta(x-y)
\\
&&
g^{\alpha\beta}({\bf v}) =E^\gamma({\bf v}) c^{\alpha\beta}_\gamma({\bf v}), \quad \Gamma^{\alpha\beta}_\gamma({\bf v}) =\left( \frac12 -\mu_\beta\right) c^{\alpha\beta}_\gamma({\bf v}).
\nn
\eeqa
Here
$$
\mu_\beta=q_\beta-\frac{d}2.
$$
The symmetric matrix $g^{\alpha\beta}({\bf v})$ is called \emph{intersection form} of the Frobenius manifold. It does not degenerate on an open dense subset in $M$. So, according to \cite{dn83} its inverse defines another flat metric on this subset. Moreover, the pair of symmetric matrices $g^{\alpha\beta}({\bf v})$ and $\eta^{\alpha\beta}$ form a so-called \emph{flat pencil} (see details in \cite{tani}). Existence of such a flat pencil is essentially equivalent to the axioms of Frobenius manifold [{\it ibid}]. Flat coordinates of the linear combination 
\beq\label{pencil}
g^{\alpha\beta}({\bf v})-\lambda \eta^{\alpha\beta}
\eeq
will play an important role in the constructions of the next section.

\begin{exam}\label{triv}
For $X={\rm pt}$ one obtains one-dimensional Frobenius manifold; all structure constants \eqref{struc} are constants. The function $\theta(v, z)$ can be found from \eqref{horizont}
\beq\label{theta0}
\theta(v, z) =\frac{e^{z\, v}-1}{z}=\sum_{p=1}^\infty \frac{v^{p+1}}{(p+1)!} z^p.
\eeq
So, in this case the principal hierarchy coincides with \eqref{hier0}.
\end{exam}

Any solution ${\bf v}={\bf v}({\bf t})$ to the principal hierarchy can be represented in implicit form as follows. Consider a function $\Phi_{{\bf t}}({\bf v})$ on the Frobenius manifold $M$ depending on the infinite vector of time variables ${\bf t}=(t^\alpha_p)$ defined as follows
\beq\label{phi}
\Phi_{{\bf t}}({\bf v})=\sum_{p\geq 0} t^\alpha_p \theta_{\alpha,p}({\bf v}).
\eeq
Choose a constant vector ${\bf t}_0=\left( {t^\alpha_p}_0\right)$ and consider a system of $n$ equations written in the vector form
\beq\label{hodo}
\nabla \Phi_{{\bf t}}({\bf v})=\nabla \Phi_{{\bf t}_0}({\bf v}).
\eeq
Suppose that the system \eqref{hodo} satisfies conditions of the implicit function theorem near the point ${\bf t}=0$. Then the solution to \eqref{hodo} can be expanded in a formal power series in $\left( t^\alpha_p\right)_{p>0}$. The resulting series satisfy equations \eqref{ph}. All solutions to the hierarchy \eqref{ph} satisfying certain genericity conditions can be obtained in such a way. Such an important statement is usually referred to as \emph{completeness} of the family of commuting Hamiltonian flows \eqref{ph}.

Choose a solution ${\bf v}({\bf t})$ in the form of \eqref{hodo} and define a function ${\mathcal F}_0({\bf t})$ by integrating the following  closed 1-form on $M$
\beq\label{gtau0}
{\mathcal F}_0({\bf t} )=\frac12 \int d\Phi_{{\bf t}-{\bf t}_0}({\bf v})\cdot  d\Phi_{{\bf t}-{\bf t}_0}({\bf v})|_{{\bf v}={\bf v}({\bf v}({\bf t})}.
\eeq
In this formula the product of 1-forms is understood in the sense of the dual Frobenius algebra structure defined on the cotangent space $T^*M$
(we refer to \cite{D1} for details about choice of integration constants in \eqref{gtau0}). The function
\beq\label{gtau00}
\tau_0({\bf t})=\exp \frac1{\epsilon^2 } {\cal F}_0({\bf t})
\eeq
is called \emph{the tau-function} of the solution \eqref{hodo} to the principal hierarchy. The solution itself can be expressed via the second logarithmic derivatives of the tau-function
\beq\label{gtau1}
v_\alpha({\bf t}) =\epsilon^2 \frac{\pal^2}{\pal x \pal t^\alpha_0} \log \tau_0({\bf t}), \quad \alpha=1, \dots, n.
\eeq
Moreover, the density of the Hamiltonian $H_{\alpha, p-1}^0$  evaluated on the solution ${\bf v}({\bf t})$ can also represented as certain second logarithmic derivatives of the tau-function
\beq\label{gtau2}
\theta_{\alpha,p}\left({\bf v}({\bf t})\right) =\epsilon^2 \frac{\pal^2}{\pal x \,\pal t^\alpha_p} \log \tau_0({\bf t}).
\eeq

For the particular choice of the shifts ${\bf t}_0$ given by
\beq\label{topsol}
{t^1_{1_0}}=1, \quad {t^\alpha_{p_0}}=0\quad \mbox{otherwise}
\eeq
the function ${\mathcal F}_0({\bf t})$ coincides with the genus zero GW total descendent potential function \eqref{fg}. For this reasons the solution ${\bf v}({\bf t})$ specified by the shifts \eqref{topsol} will be called \emph{topological solution} to the principal hierarchy. It will be denoted by ${\bf v}_{\rm top}({\bf t})$

\begin{exam} For $X={\rm pt}$ the topological solution $v=v_{\rm top}({\bf t})$ is determined by the equation
\beq\label{kw1}
x+\sum_{p=0}^\infty t_p \frac{v^p}{p!}=v
\eeq
The series expansion of this solution \eqref{kw1} reads
\beq\label{kw2}
v_{\rm top}({\bf t})=\sum_{n=1}^\infty \frac1{n} \sum_{p_1+\dots +p_n =n-1} \frac{t_{p_1}}{p_1!} \dots \frac{t_{p_n}}{p_n!}.
\eeq
The (logarithm of) the tau-function \eqref{gtau0}, \eqref{gtau00} of this solution
is given by the formula
\beq\label{kw3}
{\mathcal F}_0({\bf t})=\sum_{n=3}^\infty \frac1{n(n-1)(n-2)} \sum_{p_1+\dots +p_n =n-3} \frac{t_{p_1}}{p_1!} \dots \frac{t_{p_n}}{p_n!}.
\eeq
It coincides with the generating function of intersection numbers of $\psi$-classes on the moduli spaces $\overline{\cal M}_{0,n}$ of Riemann spheres with $n$ punctures (see more details in the next section below)
\beq\label{kw4}
{\mathcal F}_0({\bf t})=\sum_{n=3}^\infty \frac1{n!} \sum_{p_1, \dots, p_n} t_{p_1} \dots t_{p_n} \int_{\overline{\cal M}_{0,n}} \psi_1^{p_1}\dots \psi_n^{p_n}.
\eeq
\end{exam}

\subsection{Loop equations and construction of integrable hierarchy of topological type associated with a smooth projective variety with semisimple quantum cohomology}\label{sect33}\par 

Let us now explain the procedure of reconstructing of the full hierarchy written in the form of an infinite $\epsilon$-expansion starting from the Principal Hierarchy associated with a calibrated \emph{semisimple} Frobenius manifold of dimension $n$. As above we will  assume that the Frobenius manifold is realized as the quantum cohomology of a smooth projective variety $X$. Assumption of semisimplicity is a highly nontrivial restriction onto the class of varieties (see more on the problem of semisimplicity in \cite{icm}, \cite{bayer-manin}, \cite{iritani}, \cite{kawa}, \cite{teleman}, \cite{ep}, \cite{ostrov}).

\begin{lemma} Under assumption of semisimplicity the deformed flat coordinates \eqref{thetap}, \eqref{thetaz} are entire functions of the variable $z\in \mathbb C$.
\end{lemma}

We will consider $\left(\theta_1({\bf v}; z), \dots, \theta_n({\bf v}; z)\right)$ as a $H_*(X)$-valued function on $M\times \mathbb C$.

Define \emph{twisted periods} $p_\alpha^{(\nu)}({\bf v}; \lambda)$ by a Laplace-type integral
\beq\label{period}
\left( p_1^{(\nu)}({\bf v}; \lambda), \dots,  p_n^{(\nu)}({\bf v}; \lambda)\right) := \int_0^\infty\frac{dz}{z^{\frac12+\nu}} e^{-\lambda z} \left(\theta_1({\bf v}; z), \dots, \theta_n({\bf v}; z)\right) z^\mu z^{c_1(X)}.
\eeq
Here
$$
z^{c_1(X)}=1+\frac{c_1(X)}{1!} \log z+\frac{c_1^2(X)}{2!}\log^2z+\dots+\frac{c_1^d(X)}{d!}\log^dz
$$
is a polynomial in $\log z$.
The obvious formula
\beq\label{gamma}
\int_0^\infty e^{-s} s^{x-1} \log^ks\, ds=\Gamma^{(k)}(x), \quad -x\not\in \mathbb Z.
\eeq
can be used in order to give sense to the integral \eqref{period} for ${\rm Re}\, \nu <<0$. The definition can be extended to all complex values of $\nu$ satisfying
$$
\nu\not\in {\rm Spec}\,\mu+\mathbb Z_{\geq 0} -\frac12.
$$
The resulting twisted periods are analytic functions on the universal covering of
\beq\label{covering}
M\times \mathbb C\setminus \{ \det \left( {\mathcal U}({\bf v})-\lambda\right)=0\}
\eeq
where ${\mathcal U}({\bf v})$ is the operator of quantum multiplication by the Euler vector field $E({\bf v})$, see details in \cite{almost}. If the complex dimension $d=\dim_{\mathbb C}X$ is even then the twisted periods are well defined for $\nu=0$. In this case $p_1^{(0)}({\bf v}; \lambda), \dots,  p_n^{(0)}({\bf v}; \lambda)$ are flat coordinates for the $\lambda$-dependent metric \eqref{pencil}.

Introduce matrix $G(\nu)=\left( G^{\alpha\beta}(\nu)\right)$ by
\beq\label{gnu}
G(\nu) =-\frac1{2\pi} \left( e^{\pi i c_1(X)} e^{\pi i(\mu-\nu)} +e^{-\pi i c_1(X)} e^{-\pi i(\mu-\nu)} \right)\eta^{-1}
\eeq
and define coefficients $A_k^\gamma=A_k^\gamma\left( {\bf v}; {\bf v}_x, \dots, {\bf v}^{(k)}; \lambda\right)$, $B_k^\gamma=B_k^\gamma\left(  {\bf v}; {\bf v}_x, \dots, {\bf v}^{(k+2)}; \lambda\right)$, $C_{kl}^{\gamma\rho}=C_{kl}^{\gamma\rho}\left( {\bf v}; {\bf v}_x, \dots, {\bf v}^{({\rm max}\, (k, l)+1)}; \lambda\right)$, $k, \, l=0, \, 1, \, 2, \dots$, by the following formulae
\eqa\label{abc}
&&
A_k^\gamma=\pal_x^k \left( (E({\bf v})-\lambda e)^{-1}\right)^\gamma+\lim_{\nu\to 0} \sum_{j=1}^k \left( \begin{array}{c}k \\ j\end{array}\right) \pal_x^{j-1} \nabla_1 p_\alpha^{(-\nu)}({\bf v}; \lambda) G^{\alpha\beta}(\nu) \pal_x ^{k-j+1} \nabla^\gamma  p_\beta^{(\nu)}({\bf v}; \lambda)
\nn\\
&&
B_k^\gamma=\frac12 \pal_x^{k+1} \left[ \left(\lim_{\nu\to 0} G^{\alpha\beta}(\nu)\pal_\lambda\,\nabla p_\alpha^{(-\nu)}({\bf v}; \lambda) \cdot \pal_\lambda\, \nabla  p_\beta^{(\nu)}({\bf v}; \lambda)\right)\cdot {\bf v}_x\right]^\gamma 
\\
&&
\nn\\
&&
C_{kl}^{\gamma\rho} =\lim_{\nu\to 0} \pal_x^{k+1} \nabla^\gamma p_\alpha^{(-\nu)}({\bf v}; \lambda) G^{\alpha\beta}(\nu) \, \pal_x^{l+1}\nabla^\rho p_\beta^{(\nu)}({\bf v}; \lambda).
\nn
\eeqa
In these formulae all Greek indices take their values from $1$ to $n$; summation over repeated Greek indices is always assumed. The product and inversion of vector fields on $M$ in the expressions for $B_k^\gamma$ and  $A_k^\gamma$ are understood in the sense of the quantum multiplication on $TM$.

\begin{theorem} {\rm 1.} The coefficients \eqref{abc} are well defined. 

\noindent {\rm 2.} They are rational functions in $\lambda$ vanishing at $\lambda=\infty$ having poles at $\lambda\in {\rm Spec}\,{\mathcal U}({\bf v})$.

\noindent {\rm 3.} Introduce $\epsilon$-dependent linear differential operators ${\mathcal L}_m$, $m\geq -1$ acting on functions of jet variables
$$
v^{\alpha,k}:= \pal_x^k v^\alpha, \quad \alpha=1, \dots, n, \quad k=0, \, 1, \, 2, \dots  
$$
by the generating series
\eqa\label{virafull}
&&
\sum_{m\geq -1}\frac{{\mathcal L}_m}{\lambda^{m+2}} =
\\
&&
=\sum_{k\geq 0} \left( A_k^\gamma -\epsilon^2 B_k^\gamma \right)\frac{\pal}{\pal v^{\gamma, k}}-\frac{\epsilon^2}2 \sum_{k, \, l} C_{kl}^{\gamma\rho}\frac{\pal^2}{\pal v^{\gamma, k} \pal v^{\rho, l}}+\frac1{16} \tr \left( {\mathcal U}({\bf v})-\lambda\right)^{-2} -\frac14 \tr \left[ \left( {\mathcal U}({\bf v})-\lambda\right)^{-1} \mu\right]^2.
\nn
\eeqa
In this formula the operator $\mu:H^*(X) \to H^*(X)$ is defined by
\beq\label{mu}
\mu=\frac{\deg -d}2.
\eeq
These operators satisfy commutation relations
\beq\label{virfull}
[{\mathcal L}_i, {\mathcal L}_j] = (i-j) {\mathcal L}_{i+j}, \quad i, \, j\geq -1.
\eeq

\noindent {\rm 4.} Consider a system of the so-called \emph{loop equations} 
\beq\label{vc}
{\mathcal L}_k e^{\sum_{g\geq 1} \epsilon^{2g-2} {\mathcal F}_g \left({\bf v}; {\bf v}_x, {\bf v}_{xx}, \dots, {\bf v}^{(3g-2)}\right)}=0, \quad k\geq -1
\eeq
for functions ${\mathcal F}_1({\bf v}; {\bf v}_x)$, ${\mathcal F}_2 ({\bf v}; {\bf v}_x, {\bf v}_{xx}, {\bf v}_{xxx}, {\bf v}_{xxxx})$ etc.
For any smooth projective variety $X$ with semisimple quantum cohomology the system of equations \eqref{vc} has a unique, up to additive constants ${\mathcal F}_g \mapsto {\mathcal F}_g+c_g$, solution.
\end{theorem}

The proof can be derived from results of \cite{DZ}.

Equations \eqref{vc} yield a linear inhomogeneous system
\beq\label{ff1}
A_0^\gamma \frac{\pal {\mathcal F}_1}{\pal v^\gamma} + A_1^\gamma \frac{\pal{\mathcal F}_1}{\pal v^\gamma_x}=-\frac1{16} \tr \left( {\mathcal U}({\bf v})-\lambda\right)^{-2} +\frac14 \tr \left[ \left( {\mathcal U}({\bf v})-\lambda\right)^{-1} \mu\right]^2
\eeq
 for the first derivatives of the function ${\mathcal F}_1={\mathcal F}_1({\bf v}; {\bf v}_x)$ along with a recursion relation
\beq\label{ffg}
\sum_{k=0}^{3g-2} A_k^\gamma \frac{\pal {\cal F}_g}{\pal v^{\gamma,k}}=\sum_{k=0}^{3g-5} B_k^\gamma  \frac{\pal {\cal F}_{g-1}}{\pal v^{\gamma,k}}+\frac12\sum_{k, \, l=0}^{3g-5} C_{kl}^{\gamma \rho} \left( \frac{\pal^2 {\cal F}_{g-1}}{\pal v^{\gamma,k} \pal v^{\rho,l}} +\sum_{h=1}^{g-1} \frac{\pal {\cal F}_h}{\pal v^{\gamma,k}} \frac{\pal {\cal F}_{g-h}}{\pal v^{\rho,l}}\right)
\eeq
for $g\geq 2$.

The above construction of the Virasoro operators ${\cal L}_i$ is closely related to a Sugawara-type construction of \cite{DZ}. Let us briefly summarize the latter.

Introduce the Heisenberg algebra with the generators
$$
a_{\alpha,p}, ~~\alpha=1, \dots, n, ~~p\in {\mathbb Z}+{1\over 2}
$$
and the commutation relations
\beq\label{heis}
[a_{\alpha,p},a_{\beta,q}]=(-1)^{p-{1\over 2}} \eta_{\alpha\beta} \delta_{p+q, 0}.
\eeq
Introduce the row vectors
$$
{\bf a}_p = (a_{1,p}, \dots, a_{n,p})
$$
and their generating function
\beq\label{az}
{\bf a}(z)=\sum {\bf a}_p z^p.
\eeq
Define
\beq
\phi_\alpha^{(\nu)}(\lambda)=\left( \int_0^\infty {dz\over z^{1-\nu}} \,
e^{-\lambda\, z} {\bf
a}(z) z^{\mu} z^{c_1(X)}\right)_\alpha, ~~\alpha=1, \dots, n.
\eeq
Put
\beq\label{regular-T}
T^{(\nu)}(\lambda) =\sum_{m\in {\bf Z}} {L_m^{(\nu)}\over \lambda^{m+2}}
=-{1\over 2}
: \pal_\lambda \phi^{(-\nu)}_\alpha \, G^{\alpha\beta}(\nu) \pal_\lambda
\phi_\beta^{(\nu)} \, : + {1\over 4\, \lambda^2} {\rm tr}\,
\left( {1\over 4} - \mu^2\right)
\eeq
where the normal ordering is defined by
\eqa
&&:a_{\alpha,p} a_{\beta,q}: = a_{\beta,q} a_{\alpha,p} \quad
{\rm if} ~q<0, ~p>0,
\nn\\
&&
:a_{\alpha,p} a_{\beta,q}:= a_{\alpha,p} a_{\beta,q}\quad
{\rm otherwise}.\nn
\eeqa
Here $\nu$ is a complex parameter. 

\begin{lemma} {\rm \cite{DZ}} \label{lm34} Let $k$ be the minimal positive integer such that $c_1(X)^k=0$. Then
there exist the limits
\eqa
&&
L_m:= \lim_{\nu\to 0} L_m^{(\nu)}, \quad m\geq -1
\nn\\
&&
L_m:= \lim_{\nu\to 0} \nu^k L_m^{(\nu)}, \quad m<-1.
\nn\\
\eeqa
These operators satisfy the following commutation relations
\eqa
&&
[L_i, L_j] =0, ~~i, j <-1, ~{\rm or}~ i+j\geq -1, ~{\rm but} ~(i+1)(j+1)<0,
\nn\\
&&
[L_i, L_j]=(i-j) L_{i+j}, ~~i+j<-1, ~(i+1)\, (j+1)<0, ~~{\rm or} ~i, j\geq -1.
\nn
\eeqa
\end{lemma}

A natural representation of the Heisenberg algebra (\ref{heis}) is obtained as follows
\eqa
&&
a_{\alpha,p} =\epsilon{\pal\over \pal t^\alpha_{ p-{1\over 2}}}, ~~p>0,\nn\\
&&
a_{\alpha,p} = \epsilon^{-1}(-1)^{p+{1\over 2}} \eta_{\alpha\beta} t^\beta_{-p-{1\over 2}}, ~~p<0.\nn\\
\eeqa
In this representation the operators $L_m$ for $m\geq -1$ become linear second order
differential operators
\eqa
&&\hskip -1.2truecm
L_m = L_m(\epsilon^{-1}{\bf t}, \epsilon{\pal/\pal{\bf t}})\\
&&\hskip -1.2truecm
=\epsilon^2 \sum a_m^{\alpha,p; \beta,q}
{\pal^2\over \pal t^{\alpha,p} \pal t^{\beta,q}}
+ \sum {b_m}^{\alpha,p}_{\beta,q}\,  t^{\beta,q} {\pal\over \pal
t^{\alpha,p}}
+ \epsilon^{-2} c^m_{\alpha,p; \beta,q} \, t^{\alpha,p}\,  t^{\beta,q}
+{1\over 4} \delta_{m,0} \tr \, \left( {1\over 4} -{\mu}^2\right)\nn
\label{2-11-32}
\eeqa
for some constant coefficients $a_m^{\alpha,p; \beta,q}$,
${b_m}^{\alpha,p}_{\beta,q}$,
$c^m_{\alpha,p; \beta,q}$ depending on $m\geq -1$ and on
the spectrum of $\mu$ and on the first Chern class $c_1(X)$.

We are now ready to complete the construction of integrable hierarchy of topological type associated with a smooth projective variety with semisimple quantum cohomology. Solve the system of Virasoro constraints and then apply the substitution
\beq\label{subs}
v_\alpha\mapsto u_\alpha=v_\alpha+\sum_{g\geq 1}\epsilon^{2g}\frac{\partial^2 {\mathcal F}_g\left({\bf v}; {\bf v}_x, {\bf v}_{xx}, \dots, {\bf v}^{(3g-2)}\right)}{\partial x \,\partial t^\alpha_0}, \quad\alpha=1, \dots, n
\eeq
to the principal hierarchy  \eqref{ph}. We arrive at the following main construction \cite{DZ}.

\begin{theorem} \label{thm35} 1) The resulting hierarchy
\beq\label{hiertt1}
\frac{\pal u_\alpha}{
\pal t^\beta_p} =\frac{\pal v_\alpha}{\pal t^\beta_p} +\sum_{g\geq 1}\epsilon^{2g}\frac{\partial^3 {\mathcal F}_g\left({\bf v}; {\bf v}_x, {\bf v}_{xx}, \dots, {\bf v}^{(3g-2)}\right)}{\partial x \,\partial t^\alpha_0\, \pal t^\beta_p}, \quad \alpha, ~\, \beta=1, \dots, n, \quad p\geq 0
\eeq
admits a hamiltonian description
\beq\label{hiertt2}
\frac{\pal u_\alpha}{
\pal t^\beta_p} =P_{\alpha\gamma} \frac{\delta H_{\beta,p}}{\delta u_\gamma(x)}
%= \tilde P_{\alpha\gamma} \frac{\delta \tilde H_{\beta,p}}{\delta u_\gamma(x)}
\eeq
with the first Poisson bracket given by the operator
\beq\label{hiertt3}
P_{\alpha\beta}=\eta_{\lambda\mu}\,{{\rm L}}_\alpha^\lambda \circ \pal_x \circ {{\rm L}^*}^\mu_\beta
\eeq
where the matrix-valued operator ${\rm L}_\alpha^\beta$ is the linearization of the substitution \eqref{subs},
\beq\label{hiertt4}
{\rm L}_\alpha^\beta=\delta_\alpha^\beta+\sum_{k\geq 0} \frac{\pal}{\pal u_\beta^{(k)}}\left( \sum_{g\geq 1}\epsilon^{2g}\frac{\partial^2 {\mathcal F}_g\left({\bf v}; {\bf v}_x, {\bf v}_{xx}, \dots, {\bf v}^{(3g-2)}\right)}{\partial x \,\partial t^\alpha_0}\right) \frac{\pal^k}{\pal x^k},
\eeq
${{\rm L}^*}_\alpha^\beta$ is the formal adjoint operator to ${\rm L}_\alpha^\beta$.  The Hamiltonians of the hierarchy with respect to the first Poisson structure have the form
\eqa\label{hiertt5}
&&
H_{\beta,p} =\int h_{\beta,p}({\bf u};{\bf u}_x, {\bf u}_{xx}, \dots; \epsilon)\, dx 
\nn\\
&&
\\
&&
h_{\beta,p}({\bf u};{\bf u}_x, {\bf u}_{xx}, \dots; \epsilon)=\theta_{\beta, p+1}({\bf v}) +\sum_{g\geq 1}\epsilon^{2g}\frac{\partial^2 {\mathcal F}_g\left({\bf v}; {\bf v}_x, {\bf v}_{xx}, \dots, {\bf v}^{(3g-2)}\right)}{\partial x \,\partial t^\beta_{p+1}}.
\nn
\eeqa
The functions $h_{\beta, -1}=u_\beta$ are densities of Casimirs of the first Poisson bracket. The Hamiltonian densities \eqref{hiertt5} satisfy the tau-symmetry equations
\beq\label{hiertt6}
\frac{\pal h_{\alpha, p-1}}{\pal t^\beta_q} =\frac{\pal h_{\beta, q-1}}{\pal t^\alpha_p}, \quad \alpha, \, \beta=1, \dots, n, \quad p, \, q=0, \, 1, \, 2, \dots .
\eeq

2) The flows \eqref{hiertt2} are also bihamiltonian as they preserve another Poisson bracket defined by the operator $\tilde P_{\alpha\beta}$ obtained by a similar transformation applied to \eqref{vtor}. The brackets $P_{\alpha\beta}$ and $\tilde P_{\alpha\beta}$ are compatible.

3) Given a solution ${\bf v}({\bf t})$ of the Principal hierarchy, the function ${\bf u}({\bf t};\epsilon)$ defined by \eqref{subs} satisfies the hierarchy \eqref{hiertt2}. The tau-function of this solution reads
\eqa\label{hiertt7}
&&
\tau({\bf t}; \epsilon) =\exp \left[ \frac1{\epsilon^2} {\mathcal F}_0({\bf t}) +\sum_{g\geq 1}\epsilon^{2g-2} {\mathcal F}_g\left({\bf v({\bf t})}; {\bf v}_x({\bf t}), {\bf v}_{xx}({\bf t}), \dots, {\bf v}^{(3g-2)}({\bf t})\right)\right]
\nn\\
&&
\\
&&
u_\alpha({\bf t};\epsilon) =\epsilon^2 \frac{\pal^2 \log\tau ({\bf t}; \epsilon)}{\pal x\, \pal t^\alpha_0}, \quad \alpha=1, \dots, n.
\nn
\eeqa
Here the genus zero tau-function is defined by \eqref{gtau0}, \eqref{gtau00}.

4) For $m\geq -1$ linear action of Virasoro operators $L_m$ defined in Lemma \ref{lm34} on tau-functions generates infinitesimal symmetries of the hierarchy \eqref{hiertt2}, i.e., given a tau-function $\tau=\tau({\bf t};\epsilon)$ of a solution to the hierarchy \eqref{hiertt2},  then for an arbitrary small parameter $\delta$ the functions
$$
u_\alpha({\bf t};\epsilon;\delta) =\epsilon^2 \frac{\pal^2 \log\left(\tau +\delta \cdot L_m \tau\right)}{\pal x\, \pal t^\alpha_0}
$$
for any $m\geq -1$ satisfy, modulo corrections of order ${\mathcal O}(\delta^2)$, the same equations of the hierarchy \eqref{hiertt2} .

5) Denote ${\bf u}_{\rm top} ({\bf t};\epsilon)$ the solution to the hierarchy \eqref{hiertt2} obtained from the topological solution ${\bf v}_{\rm top}({\bf t})$ to the principal hierarchy (see above). Let $\tau_{\rm top}({\bf t};\epsilon)$ be the tau-function \eqref{hiertt7} of ${\bf u}_{\rm top}({\bf t};\epsilon)$. It satisfies the \emph{Virasoro constraints}
\beq\label{hiertt8}
L_m \left( \epsilon^{-1} ({\bf t}-{\bf t}_0\right), \epsilon\, \pal/\pal {\bf t}) \tau_{\rm top}({\bf t};\epsilon)=0, \quad m\geq -1.
\eeq
\end{theorem}

\begin{cor} Assuming validity of the Virasoro conjecture \cite{ehx} for the variety $X$, the logarithm of the topological tau-function $\tau_{\rm top}({\bf t};\epsilon)$ coincides with the total GW potential \eqref{total}.
\end{cor}

\begin{exam} For $X={\rm pt}$ the twisted periods are
$$
p^{(\nu)}(v;\lambda)=\Gamma\left(-\frac12-\nu\right) \left( \lambda-v\right)^{\nu+\frac12}
$$
and the $1\times 1$ matrix \eqref{gnu} reads
$$
G(\nu)=-\frac1{\pi}\cos\pi \nu.
$$
One can easily see that the general formulae \eqref{abc}--\eqref{ffg} reduce to \eqref{jvir1}--\eqref{virec}. Actually in this case one can replace in \eqref{abc} from the very beginning $p^{(\pm \nu)}$ by $p^{(0)}=\sqrt{v-\lambda}$ (up to a constant factor). Observe that $\sqrt{v-\lambda}$ coincides with the flat coordinate of the flat pencil of metrics \eqref{pencil}, i.e., of the $\lambda$-dependent metric 
$$
\frac{dv^2}{v-\lambda}
$$
obtained by inversion of \eqref{pencil}. The topological tau-function $\tau_{\rm top}({\bf t};\epsilon)$ coincides with the Witten--Kontsevich tau-function of the KdV hierarchy. It gives the generating function of intersection numbers of $\psi$-classes on the Deligne--Mumford moduli spaces $\overline{\mathcal M}_{g,n}$.

The Virasoro symmetries of the KdV hierarchy are generated by  the following linear operators
\eqa\label{kdv-vir}
&&
L_m={\epsilon^2\over 2} \sum_{k+l=m-1} {(2k+1)!!\,(2l+1)!!\over 2^{m+1}}
{\pal^2\over \pal t_k \pal t_l}\nn\\
&&\qquad + \sum_{k\geq 0} {(2k+2m+1)!!\over 2^{m+1}
(2k-1)!!}\, t_k\,{\pal\over\pal t_{k+m}}+{1\over 16} \delta_{m,0},
\quad m\geq 0,
\nn\\
&&
L_{-1}=\sum_{k\geq 1} t_k{\pal\over \pal t_{k-1}}
+{1\over 2\epsilon^2} t_0^2
\nn
\eeqa
The Witten--Kontsevich tau-function $\tau=\tau_{\rm top}({\bf t})$is uniquely specified by the system of linear equations
\eqa\label{kw-constraints}
&&
L_m \tau = \frac{(2m+3)!!}{2^{m+1}} \pal_{m+1}\tau, ~~m\geq 0
\nn\\
&&
L_{-1} \tau =\pal_0 \tau.\nn
\eeqa

\end{exam}

\begin{exam} \label{cp1} For $X={\bf P}^1$ the basis is $\gamma_1=1$, $\gamma_2=\omega \in H^2({\bf P}^1)$,
$$
\int_{{\bf P}^1}\omega=1.
$$
In order to simplify notations we will redenote 
$$
v^1\to v, \quad v^2\to u
$$
the corresponding flat coordinates on the two-dimensional Frobenius manifold $M=H^*({\bf P}^1)$. So the notation
$$
{\bf v} =v+u\,\omega\in M
$$
will still be used for a generic point in the Frobenius manifold.
The potential of the Frobenius structure, the Gram matrix of the flat metric, the unity, and the Euler vector field read
$$
F=\frac12 v^2 u+e^u,\quad \eta=\left(\begin{array}{cc}0 & 1\\ 1 & 0\end{array}\right), \quad e=\frac{\pal}{\pal v}, \quad E=v\frac{\pal}{\pal v}+2\frac{\pal}{\pal u}.
$$
Thus the operator of multiplication by $E$ has the following matrix
$$
{\cal U}({\bf v})=\left(\begin{array}{cc} v & 2 e^u\\ 2 & v\end{array}\right).
$$
The deformed flat coordinates can be expressed via modified Bessel functions
\eqa
&&\theta_1({\bf v};z)=\sum_{p\geq 0} \theta_{1,p}(v,u) z^p=-2\,e^{z v}\left(K_0(2 z e^{\frac12\,u})
+(\log{z}+\gamma) I_0(2 z e^{\frac12\,u})\right)\nn\\
\label{theta-toda}\\
&&\theta_2({\bf v};z)=\sum_{p\geq 0} \theta_{2,p}(v,u) z^p=z^{-1}\left[e^{z v}\,I_0(2 z e^{\frac12\,u})-1\right]\nn
\eeqa
Here $\gamma$ is the Euler--Mascheroni constant. The time variables of the Principal Hierarchy will be redenoted
$$
t^1_k\to t_k, \quad t^2_k \to s_k.
$$
So
\eqa\label{princp1} 
&&
\frac{\pal v}{\pal t_k} =\frac{\pal}{\pal x}\frac{\pal \theta_{1,k+1}(v,u)}{\pal u}, \quad \frac{\pal u}{\pal t_k} =\frac{\pal}{\pal x} \frac{\pal \theta_{1,k+1}(v,u)}{\pal v}
\nn\\
&&
\frac{\pal v}{\pal s_k} =\frac{\pal}{\pal x} \frac{\pal \theta_{2,k+1}(v,u)}{\pal u}, \quad \frac{\pal u}{\pal s_k} =\frac{\pal}{\pal x} \frac{\pal \theta_{2,k+1}(v,u)}{\pal v}.
\nn
\eeqa
In particular,
\beq\label{s0}
\frac{\pal v}{\pal s_0} = e^u u_x, \quad \frac{\pal u}{\pal s_0} =v_x.
\eeq
A basis of twisted periods, after a suitable linear combination, can be expressed via Legendre functions\footnote{Also called Ferrers functions.} of the 1st and 2nd kind
\eqa\label{pernu}
&&
p_1^{(\nu)}({\bf v}; \lambda) = \frac1{\nu} e^{-\frac{u}4} \Delta^{\frac14 +\frac{\nu}2} P_{-\frac12}^{\frac12+\nu}(w)
\nn\\
&&
\\
&&
p_2^{(\nu)}({\bf v}; \lambda) = \frac1{\nu} e^{-\frac{u}4} \Delta^{\frac14 +\frac{\nu}2} Q_{-\frac12}^{\frac12+\nu}(w)
\nn
\eeqa
where
$$
\Delta=-\det ( {\cal U}({\bf v})-\lambda)=4 \,e^u -(v-\lambda)^2, \quad w=\frac12 e^{-\frac{u}2} (\lambda-v).
$$
The matrix \eqref{gnu} in this basis becomes
\beq\label{gnu1}
G(\nu)=\left(\begin{array}{rr} \frac{\pi}2 \cos\pi\nu & -\sin\pi\nu\\
\\
\sin\pi\nu & \frac2{\pi} \cos\pi\nu\end{array}\right).
\eeq
For the coefficients \eqref{abc} one easily obtains
\eqa
&&
A_0^1=\frac{\lambda-v}{\Delta}, \quad A_0^2 =\frac2{\Delta}
\nn\\
&&
A_1^1=-\frac{8 e^u +(v-\lambda)^2}{\Delta^2} v_x +\frac{6 e^u(v-\lambda)}{\Delta^2} u_x, \quad A_1^2=\frac{6(v-\lambda)}{\Delta^2} v_x -\frac{12 \,e^u}{\Delta^2} u_x.
\nn
\eeqa
Furthermore, since $\mu={\rm diag} \left( -\frac12, \frac12\right)$ we have
$$
\tr ({\cal U}({\bf v})-\lambda)^{-2}=2\frac{4 e^u+(v-\lambda)^2}{\Delta^2},\quad \tr \left[ ({\cal U}({\bf v})-\lambda)^{-1}\mu\right]^2=-\frac1{2\Delta}.
$$
So the equation \eqref{ff1} yields
\beq\label{fcp1}
{\cal F}_1 = \frac1{24}\left[ \log \left(v_x^2 -e^u u_x^2\right) - u\right].
\eeq
This formula was derived in \cite{dz98} from topological considerations. 

A somewhat more lengthy computation allows one to also derive higher genus terms. For example (cf. \cite{egx}),
\eqa\label{fcp1_2}
&&
5760 \,{\mathcal F}_2 = 
\nn\\
&&
=-\frac{q^2}{D^4} \left[ 512 u_x^3 v_x v_{xx}^3 +384 q u_x^3 v_{xx} (u_x^2+2 u_{xx}) (u_x^2 v_x+2 u_{xx} v_x -2 u_x v_{xx})-64 q^2 u_x^4 (u_x^2 +2 u_{xx})^3\right]
\nn\\
&&
-\frac{q}{D^3}\left[ 256 u_x v_x v_{xx}^3 +12 q u_x \left( 28 u_x^4 v_x v_{xx} +116 u_x^2 u_{xx} v_x v_{xx} +64 u_{xx}^2 v_x v_{xx} + 28 u_x v_x u_{xxx} v_{xx} -69 u_x^3 v_{xx}^2 \right.\right.
\nn\\
&&
\left.\left.
-128 u_x u_{xx} v_{xx}^2 +14 u_x^3 v_x v_{xxx} +28 u_x v_x u_{xx} v_{xxx} -28 u_x^2 v_{xx} v_{xxx}\right)\right.
\nn\\
&&\left.
-q^2 u_x^2 (u_x^2 +2 u_{xx}) (121 u_x^4+538 u_x^2 u_{xx} +256 u_{xx}^2 +168 u_x u_{xxx})
\right]
\nn\\
&&
+\frac{q}{D^2} \left[-2\left( 42 u_x^3 v_x v_{xx} +126 u_x u_{xx} v_x v_{xx} +42 u_{xxx} v_x v_{xx} -95 u_x^2 v_{xx}^2 -96 u_{xx} v_{xx}^2+30 u_x^2 v_x v_{xxx} \right.\right.
\nn\\
&&\left.\left.
+42 u_{xx} v_x v_{xxx} -126 u_x v_{xx} v_{xxx} +20 u_x v_x v_{xxxx}\right)+q\left( 72 u_x^6+479 u_x^4 u_{xx}+626 u_x^2 u_{xx}^2 +64 u_{xx}^3 \right.\right.
\nn\\
&&\left.\left.
+224 u_x^3 u_{xxx} +252 u_x u_{xx} u_{xxx} +40 u_x^2 u_{xxxx}\right)
\right]-\frac1{D}\left[22 v_{xx}^2-24 v_x v_{xxx} +q\left( 17 u_x^4 +102 u_x^2 u_{xx}\right.\right.
\nn\\
&&\left.\left. 
+56 u_{xx}^2 +68 u_x u_{xxx} +20 u_{xxxx}\right)
\right]+7u_{xx}
\eeqa
where we denote
$$
q=e^u, \quad D=v_x^2 -e^u u_x^2.
$$

Applying the quasitriviality substitution
\eqa\label{quasicp1}
&&
u\mapsto u+\epsilon^2 \pal_x^2\Delta{\mathcal F}
\nn\\
&&
v\mapsto v+\epsilon^2 \pal_x \pal_{s_0} \Delta{\mathcal F}
\nn
\eeqa
where the $s_0$-derivatives are defined by \eqref{s0} and
$$
\Delta{\mathcal F}={\mathcal F}_1+\epsilon^2 {\mathcal F}_2+\dots
$$
one obtains the \emph{extended Toda hierarchy} \cite{cdz}, \cite{cmp-vir}. The latter can be represented in the Lax form
\eqa
&&
\epsilon\,\frac{\pal L}{\pal s_k} =\frac1{(k+1)!} \left[ (L^{k+1})_+, L\right]
\nn\\
&&
\\
&&
\epsilon\,\frac{\pal L}{\pal t_k}=\frac2{k!} \left[ L^k (\log L -c_k)_+, L\right], \quad c_k=1+\frac12+\dots+\frac1{k}.
\nn
\eeqa
Here $L$ is a degree two difference operator acting on functions on the $x$-axis by
$$
L=\Lambda + v(x)+e^{u(x)}\Lambda^{-1}
$$
where $\Lambda$ stands for the shift operator
$$
\Lambda \, f(x) =f(x+\epsilon),
$$
the symbol $(~)_+$ refers to the part of a difference operator containing only nonnegative degrees $\Lambda^k$, $k\geq 0$. We refer the reader to \cite{cdz} for the definition of logarithm of a difference operator.

The Virasoro symmetries of the extended Toda hierarchy are generated by the following linear operators
\eqa
&&
 L_m={\epsilon^2} \sum_{k=1}^{m-1} k!\, (m-k)!
{\pal^2\over \pal s_{k-1}
\pal s_{m-k-1}}\nn\\
&&+\sum_{k\geq 1} {(m+k)!\over (k-1)!}\left(
t_k {\pal\over \pal t_{ m+k}}
+ s_{k-1}{\pal\over \pal s_{m+k-1}}\right)
+2\,\sum_{k\geq 0} \alpha_m(k) t_k {\pal \over \pal s_{ m+k-1}},
\quad m>0
\nn\\
&&
L_0 =\sum_{k\geq 1} k\, \left( t_k {\pal\over \pal t_k}+s_{k-1}
{\pal\over \pal s_{k-1}}\right)+\sum_{k\geq 1} 2\, t_k {\pal\over \pal
s_{k-1}} + {1\over\epsilon^{2}}{t_0}^2,
\nn\\
&&
L_{-1}=\sum_{k\geq 1} \left(t_k {\pal\over \pal t_{ k-1}}+s_k {\pal\over \pal s_{ k-1}}\right)+{1\over
\epsilon^2 } t_0 s_0.
\nn
\eeqa
Here the integer coefficients $\alpha_m(k)$ are defined by
$$
\alpha_m(0)= m!, ~~\alpha_m(k) = {(m+k)!\over (k-1)!} \left[ \psi(k+m+1)-\psi(k)\right], ~k>0
$$
where $\psi(x)$ is the digamma function.
\end{exam}

For other smooth projective varieties $X$ with semisimple quantum cohomology very little is known about the associated integrable hierarchies of topological type. The conjectural description of the integrable hierarchy of topological type associated with the orbifold quantum cohomology of ${\bf P}^1$ with two orbifold points is given by the so-called bigraded Toda hierarchy \cite{mt}, \cite{carlet}. A somewhat more general setting of GW theory of the resolved conifold ${\mathcal O}_{{\bf P}^1}(-1)\oplus {\mathcal O}_{{\bf P}^1}(-1)$ with respect to the antidiagonal torus action has been recently studied in \cite{bcr} in connection with the Ablowitz--Ladik integrable hierarchy. A one-parameter deformation of the KdV hierarchy closely connected with the intermediate long wave equation was recently obtained in \cite{buryak} by inserting certain combinations of Hodge $\lambda$-classes into the Witten--Kontsevich generating series. Except for these examples and also Drinfeld--Sokolov hierarchies of $A\, D\, E$ type \cite{DS}, \cite{ds1}, \cite{fgb}, \cite{wu} related to computation of intersection numbers on the moduli spaces of algebraic curves with higher spin structures \cite{wit} and their generalizations \cite{fjr} other integrable hierarchies of topological type arising in the quantum cohomology setting seem to be unknown in the theory of integrable systems.

\begin{remark} In \cite{givental} a so-called \emph{total descendent potential} was associated with an arbitrary calibrated semisimple Frobenius manifold. Identifying Virasoro constraints of \cite{givental} with those constructed in \cite{DZ} (see above the explicit formulae) one can prove that the Givental's total descendent potential is equal to the logarithm of the tau-function of the topological solution to the integrable hierarchy of topological type associated with the Frobenius manifold.
\end{remark}

\subsection{On axiomatic approach to integrable hierarchies of topological type}\par

What can be said about properties of the integrable hierarchies of topological type constructed by Theorem \ref{thm35} above? Clearly, they remain integrable since they are obtained by change of dependent variables from the integrable principle hierarchy. Moreover, applying the same quasitriviality substitution to the bihamiltonian structure of the principal hierarchy one obtains the bihamiltonian structure for the resulting hierarchy of topological type. For example, we will write here the formula for the second Poisson brackets of the variables $u_1$. For $d\neq 1$ it has a Virasoro-type structure
\beq\label{cmp98}
\{ u_1(x) , u_1(y)\}_2 =\frac{1-d}2\left[ u_1(x)+u_1(y)\right] \delta'(x-y) +\frac{\epsilon^2}2 \tr\left[ \frac14 -\mu^2\right]\delta'''(x-y)+{\mathcal O}\left( \epsilon^4\right)
\eeq
within the $\epsilon^2$ approximation. 

One more property is \emph{existence of a tau-function} for the constructed hierarchy due to tau-symmetry \eqref{hiertt6}. Moreover, the full hierarchy is invariant with respect to a group of symmetries generated by linear action of Virasoro operators $L_m$, $m\geq -1$, onto tau-function.

In \cite{DZ} it was formulated a problem of classification of integrable hierarchy of bihamiltonian evolutionary partial differential equations of the form
\beq\label{ansa}
\frac{\pal{\bf u}}{\pal t^\alpha_p} =\sum_{n\geq 0}\epsilon^n {\bf K}_{n; \alpha, p} \left({\bf u}; {\bf u}_x, {\bf u}_{xx}, \dots, {\bf u}^{(n+1)}\right)
\eeq 
where ${\bf K}_{n; \alpha, p} \left({\bf u}; {\bf u}_x, {\bf u}_{xx}, \dots, {\bf u}^{(n+1)}\right)$ for any $n\geq 0$ is a polynomial in the jet variables ${\bf u}_x$, ${\bf u}_{xx}$, \dots, ${\bf u}^{(n+1)}$ of graded degree $n+1$, i.e., 
$$
{\bf K}_{n; \alpha, p} \left({\bf u}; \lambda{\bf u}_x, \lambda^2{\bf u}_{xx}, \dots, \lambda^{n+1}{\bf u}^{(n+1)}\right)=\lambda^{n+1} {\bf K}_{n; \alpha, p} \left({\bf u}; {\bf u}_x, {\bf u}_{xx}, \dots, {\bf u}^{(n+1)}\right)
$$
for any $\lambda$ satisfying the tau-symmetry condition and also invariant with respect to symmetries generated by a linear action onto tau-functions of Virasoro operators of the above form. It was shown that such a hierarchy, under an additional assumption of semisimplicitly of the linear operator ${\bf u}_x \mapsto {\bf K}_{0; \alpha, p}({\bf u}; {\bf u}_x)$ for at least one pair of indices $\alpha$, $p$ and also a certain condition of nondegeneracy  of the $\epsilon=0$ term of the first Poisson bracket, is equivalent, up to a \emph{Miura-type transformation}
\beq\label{miura}
{\bf u} \mapsto {\bf F}_0({\bf u}) +\sum_{n\geq 1} \epsilon^n {\bf F}_n\left({\bf u}; {\bf u}_x, \dots, {\bf u}^{(n)}\right)
\eeq
satisfying
$$
\det\left( \frac{\pal F^\alpha_0({\bf u})}{\pal u^\beta}\right)\neq 0,
$$
the graded degree of differential polynomials ${\bf F}_n\left({\bf u}; {\bf u}_x, \dots, {\bf u}^{(n)}\right)$ is equal to $n$, to an integrable hierarchy of topological type constructed from a semisimple Frobenius manifold by the construction of the Theorem \ref{thm35} (in this more general case the construction will depend on the choice of calibration, i.e., on the choice of a basis of horizontal  sections of the deformed flat connection $\tilde\nabla$).

In order to complete the proposed axiomatic approach to the theory of integrable hierarchies of topological type one has to fix the problem of cancellation of denominators. Namely,
like in the KdV case the functions ${\mathcal F}_g \left({\bf v}; {\bf v}_x, {\bf v}_{xx}, \dots, {\bf v}^{(3g-2)}\right)$ are not differential polynomials. Nevertheless the equations of the resulting integrable hierarchy of topological type proved to be polynomial in jet variables ${\bf v}_x$, ${\bf v}_{xx}$ etc. at every order of the $\epsilon$-expansion (only even powers of $\epsilon$ occur). Cancellation of denominators in the equations of the hierarchy, in their Hamiltonians and in the \emph{first} Poisson bracket was proved by A.Buryak, H.Posthuma and S.Shadrin \cite{bps}. So the last problem to be fixed in order to complete the proposed axiomatic formulation of the theory of integrable hierarchies of topological type is to also prove polynomiality of the second Poisson bracket. It would also be very interesting to find a Lax representation (see \eqref{lax} above) for the integrable hierarchies of topological type. First steps in this direction have been done in \cite{casha}.

\setcounter{equation}{0}
\setcounter{theorem}{0}
\section{Integrable hierarchies at the degree zero approximation}\label{sect4}\par

In this section we consider much simpler situation of intersection theory on moduli stacks of stable maps of degree zero. In this case, under certain not very restrictive assumption about a smooth projective variety $X$ we will construct explicitly the associated integrable hierarchy involved in description of invariants of all genera. In this section we will assume that $X$ is a Fano variety with vanishing $H^{\rm odd}(X, {\mathbb C})$ of the complex dimension
\beq\label{d4}
d\geq 4.
\eeq

Holomorphic maps to $X$ of degree $\beta=0$ are just maps to a point
$C_g \to {\rm pt}\in X$. So, at a first glance it looks like the moduli space $\overline{M}_{g,m} (X,\beta=0)$ splits into a Cartesian product of the Deligne--Mumford moduli space $\overline{\mathcal M}_{g,m}$ of stable algebraic curves of genus $g$ with $m$ punctures and the variety $X$ itself. The situation becomes more delicate if we look at the virtual fundamental class of $\overline{M}_{g,m} (X,\beta=0)$.

%\subsection{On invariants of degree zero and genus zero}\par

Let us begin again with genus zero. In this case the moduli space $\overline{M}_{0,m} (X,\beta=0)$ is smooth, so, indeed,
$$
\overline{M}_{0,m} (X,\beta=0)=\overline{\mathcal M}_{0,m}\times X.
$$
So
\beq\label{gb0}
 \left\langle \tau_{p_1}(\gamma_{\alpha_1})\dots \tau_{p_m}(\gamma_{\alpha_m})\right\rangle_{g=0, m, \beta=0}=\int_{\overline{\mathcal M}_{0,m}} \psi_1^{p_1} \dots \psi_m^{p_m} \int_X \gamma_{\alpha_1} \dots \gamma_{\alpha_m}.
\eeq
The Frobenius manifold $M$ in this case is trivial: the algebras on the tangent spaces to $M$ are canonically isomorphic to the cohomology ring $H^*(X)$. The Principal Hierarchy for such a Frobenius manifold takes the form
\beq\label{phb0}
\frac{\partial{\bf v}}{\partial t^\alpha_p}=\frac{\partial}{\partial x} \left(\gamma_\alpha\cdot \frac{{\bf v}^{p+1}}{(p+1)!}\right), \quad \alpha=1, \dots, n=\dim H^*(X), \quad p\geq 0.
\eeq
The bihamiltonian structure of this hierarchy obtained from the general prescription of the theory of Frobenius manifolds can be written in the following form
\eqa\label{Biham0}
&&
\left\{ \langle a, {\bf v}(x)\rangle, \langle b , {\bf v}(y)\rangle\right\}_1=\langle a, b\rangle \delta'(x-y)
\nn\\
&&
\\
&&
\left\{ \langle a, {\bf v}(x)\rangle, \langle b , {\bf v}(y)\rangle\right\}_2=
\nn\\
&&
=\left[\left\langle \left( \frac{a}2 +\mu(a)\right)\cdot b, {\bf v}(x) \right\rangle + \left\langle a\cdot \left( \frac{b}2 +\mu(b)\right) , {\bf v}(y)\right\rangle+\left\langle a\cdot b, c_1(X)\right\rangle\right]\delta'(x-y).
\nn
\eeqa
Here $a$, $b\in H^*(X)$ are arbitrary cohomology classes. The genus zero tau-function is given by the formula
\beq\label{Tau00}
\epsilon^2 \log\tau_0=\sum_{m\geq 3} \frac1{m(m-1)(m-2)}\sum_{p_1,\dots,p_m} \int_X \frac{{\bf t}_{p_1}}{p_1!}
\dots \frac{{\bf t}_{p_m}}{p_m!}
\eeq
where, like above, we consider cohomology-valued time variables ${\bf t}_p= t^\alpha_p \gamma_\alpha\in H^*(X)$.

Let us now proceed to the description of higher genera.

\begin{theorem} For $\dim_{\mathbb C}X\geq 4$ the total Gromov--Witten potential of degree zero is (log of) a tau-function of the following integrable hierarchy
\beq\label{htt0}
\frac{\partial {\bf u}}{\partial t^\alpha_p} =\frac{\partial}{\partial x} \left( \gamma_\alpha\cdot \left[ \frac{{\bf u}^{p+1}}{(p+1)!}+\frac{\epsilon^2}{24} c_d \cdot \left( 2\frac{{\bf u}^{p-1}}{(p-1)!} {\bf u}_{xx} +\frac{{\bf u}^{p-2}}{(p-2)!} {\bf u}_x^2\right) - \frac{\epsilon^2}{24} c_{d-1} \cdot \frac{{\bf u}^{p-1}}{(p-1)!} {\bf u}_x^2\right]\right) 
\eeq
where $c_1=c_1(X)$ and $c_{d-1}=c_{d-1}(X)$ are Chern classes of the tangent bundle of $X$.
\end{theorem}

The \emph{proof} is based on the following

\begin{lemma} For any smooth projective variety $X$ the genus one and degree zero Gromov--Witten potential is given by the following formula
\beq\label{gw10}
{\mathcal F}_1 = \frac1{24} \langle c_d(X), \log {\bf v}_x\rangle -\frac1{24} \langle c_{d-1}(X), {\bf v}\rangle
\eeq
where ${\bf v}={\bf v}({\bf t})$ is the solution
\beq\label{KW2}
{\bf v}=\sum_{n=1}^\infty \frac1{n} \sum_{p_1+\dots +p_n =n-1} \frac{{\bf t}_{p_1}}{p_1!} \dots \frac{{\bf t}_{p_n}}{p_n!}
\eeq
to the hierarchy \eqref{phb0} (cf. formula \eqref{Tau00} for the tau function of this solution).
\end{lemma} 

\pf For $\beta=0$ the dimension of the ``na\"{\i}ve" moduli space $\overline{\mathcal M}_{g,m}\times X$ is equal to $3g-3+m+d$ while the dimension \eqref{dimvirt} of the virtual fundamental class is smaller 
\beq\label{virtdim}
\dim\left[\overline{M}_{g,m} (X,\beta=0)\right]^{\rm virt} = 3g-3+m+d-g\, d.
\eeq
The discrepancy comes from the \emph{obstruction bundle} (see details in the book \cite{manin}). The following formula for the degree zero invariants of any genus is due to M.Kontsevich and Yu.I.Manin \cite{km97}
\beq\label{km97}
\left\langle \tau_{p_1}(\gamma_{\alpha_1})\dots \tau_{p_m}(\gamma_{\alpha_m})\right\rangle_{g, m, \beta=0}=\int_{\overline{\mathcal M}_{g,m}\times X} \psi_1^{p_1} \dots \psi_m^{p_m}e \left( {\mathcal E}^* \boxtimes T_X\right) \gamma_{\alpha_1} \dots \gamma_{\alpha_m}.
\eeq
Here $e \left( {\mathcal E}^* \boxtimes T_X\right)$ is the Euler class of the obstruction bundle over $\overline{\mathcal M}_{g,m}\times X$. The first factor ${\mathcal E}^*$ is the dual to the Hodge bundle over $ \overline{\mathcal M}_{g,m}$ whose fiber coincides with the space of holomorphic differentials on the curve $C_g$; the second factor is the tangent bundle of $X$.
%$$
% \begin{array}{c} {\mathcal E}  \\\Big\downarrow\vcenter{%
%\rlap{$\scriptstyle{\Gamma(\Omega(C_g))}$}}   \\ \overline{\mathcal M}_{g,m} \end{array}
%$$
Since $d\geq 4$ the virtual dimension \eqref{virtdim} is less than $m$ for $g\geq 2$. Hence \cite{manin}
\beq\label{vanish}
\left\langle \tau_{p_1}(\gamma_{\alpha_1})\dots \tau_{p_m}(\gamma_{\alpha_m})\right\rangle_{g, m, \beta=0}=0\quad \mbox{if}\quad g\geq 2.
\eeq
It remains to compute the Euler class of the obstruction bundle for $g=1$. An easy calculation yields
\beq\label{euler1}
e \left( {\mathcal E}^* \boxtimes T_X\right) =c_d\left( {\mathcal E}^* \boxtimes T_X\right) = 1\boxtimes c_d(X)- \lambda_1\boxtimes c_{d-1}(X).
\eeq
Here we use the standard notation
$$
\lambda_i=c_i({\mathcal E})
$$
for Chern classes of the Hodge bundle. Thus
\eqa\label{gp0}
&&
\left\langle \tau_{p_1}(\gamma_{\alpha_1})\dots \tau_{p_m}(\gamma_{\alpha_m})\right\rangle_{g=1, m, \beta=0}=
\\
&&
=\int_{\overline{\mathcal M}_{1,m}} \psi_1^{p_1} \dots \psi_m^{p_m} \int_X c_d(X) \gamma_{\alpha_1} \dots \gamma_{\alpha_m}-\int_{\overline{\mathcal M}_{1,m}} \lambda_1\psi_1^{p_1} \dots \psi_m^{p_m} \int_X c_{d-1}(X) \gamma_{\alpha_1} \dots \gamma_{\alpha_m}.
\nn
\eeqa
Let us begin with the first term. We know (see \cite{dw90}) that the generating function of the intersection numbers of the genus one $\psi$-classes is equal to
\beq\label{dw}
\sum_m \frac1{m!} \sum_{p_1, \dots, p_m} t_{p_1} \dots t_{p_m}\int_{\overline{\mathcal M}_{1,m}} \psi_1^{p_1} \dots \psi_m^{p_m}=\frac1{24} \log v_x({\bf t})
\eeq
(cf. formula \eqref{fen1} above) where $v({\bf t})$ is the solution \eqref{kw1}, \eqref{kw2} of the dispersionles KdV hierarchy. Thus 
\beq\label{dw1}
\sum_m \frac1{m!} \sum_{(p_1,\alpha_1), \dots, (p_m, \alpha_m)} t^{\alpha_1}_{p_1}\dots t^{\alpha_m}_{p_m}  \int_{\overline{\mathcal M}_{1,m}} \psi_1^{p_1} \dots \psi_m^{p_m} \int_X c_d(X) \gamma_{\alpha_1} \dots \gamma_{\alpha_m}=\frac1{24} \left\langle c_d(X), \log {\bf v}_x({\bf t})\right\rangle
\eeq
where
\beq\label{VT}
{\bf v}({\bf t})=\sum_{n=1}^\infty \frac1{n} \sum_{p_1+\dots +p_n =n-1} \frac{{\bf t}_{p_1}}{p_1!} \dots \frac{{\bf t}_{p_n}}{p_n!}
\eeq
is a particular solution to the hierarchy \eqref{phb0} constructed by a procedure similar to \eqref{kw2}.

Let us now proceed to the second term in \eqref{gp0}. We will use the formula
\beq\label{gp1}
\int_{\overline{\mathcal M}_{1,m}} \lambda_1\psi_1^{p_1} \dots \psi_m^{p_m}=\left\{ \begin{array}{cl}\frac1{24} \frac{(m-1)!}{p_1! \dots p_m!}, & p_1+\dots+p_m=m-1\\
\\
0, & \mbox{otherwise}\end{array}\right.
\eeq
derived by E.Getzler and R.Pandharipande \cite{gp98}. So
\beq
\sum_m\frac1{m!} \sum_{(p_1,\alpha_1), \dots, (p_m, \alpha_m)} t^{\alpha_1}_{p_1}\dots t^{\alpha_m}_{p_m}  \int_{\overline{\mathcal M}_{1,m}} \lambda_1\psi_1^{p_1} \dots \psi_m^{p_m} \int_X c_{d-1}(X) \gamma_{\alpha_1} \dots \gamma_{\alpha_m}=\frac1{24} \left\langle c_{d-1}(X),  {\bf v}({\bf t})\right\rangle.
\eeq
This completes the proof of Lemma. \epf

Since $d\geq 4$ the higher genus terms of degree zero all vanish. That is the degree zero part of the total Gromov--Witten potential is equal to
\beq\label{total0}
{\cal F} = {\cal F}_0 +\epsilon^2 {\cal F}_1.
\eeq
According to the general scheme explained in the previous sections we have to apply the substitution
\beq\label{subs0}
v_\alpha =\langle \gamma_\alpha, {\bf v}\rangle \mapsto u_\alpha =\langle \gamma_\alpha, {\bf u}\rangle =v_\alpha+\epsilon^2 \frac{\pal^2 {\mathcal F}_1}{\pal x\, \pal t^\alpha_0}
\eeq
to the dispersionless hierarchy \eqref{phb0}.
Since
$$
\frac{\pal{\bf v}}{\pal t^\alpha_0} =\gamma_\alpha\cdot {\bf v}_x,
$$
we have
$$
\frac{\pal}{\pal t^\alpha_0} \langle c_d(X), \log {\bf v}_x\rangle=\left\langle c_d(X)\cdot\gamma_\alpha, \frac{{\bf v}_{xx}}{{\bf v}_x}\right\rangle =\pal_x \langle \gamma_\alpha, c_d(X)\cdot \log {\bf v}_x\rangle.
$$
In a similar way
$$
\frac{\pal}{\pal t^\alpha_0} \langle c_{d-1}(X),  {\bf v}\rangle=\pal_x \langle \gamma_\alpha, c_{d-1}(X)\cdot {\bf v}\rangle.
$$
So the substitution \eqref{subs0} can be written in the vector form
\beq\label{subs00}
{\bf v} \mapsto {\bf u}={\bf v}+\frac{\epsilon^2}{24} c_d\cdot \left(\log{\bf v}_x\right)_{xx} -\frac{\epsilon^2}{24} c_{d-1} \cdot {\bf v}_{xx}.
\eeq
Applying this substitution to the hierarchy \eqref{phb0} one easily arrives at the equations \eqref{htt0}. \epf

One may observe similarity of eq. \eqref{htt0} with the first two terms of expansion of equations of the KdV hierarchy \eqref{kdv4}. One major difference is that the equations of the hierarchy \eqref{htt0} truncate at the order $ \epsilon^2$ while the $i$-th equation of the KdV hierarchy contains terms up to the order $\epsilon^{2i}$.

It is not difficult to apply the substitution \eqref{subs00} to the bihamiltonian structure \eqref{Biham0} in order to arrive at the bihamiltonian structure of the equations \eqref{htt0}. The resulting bihamiltonian structure reads
\eqa\label{Biham1}
&&
\left\{ \langle a, {\bf u}(x)\rangle, \langle b , {\bf u}(y)\rangle\right\}_1=\langle a, b\rangle \delta'(x-y)-\frac{\epsilon^2}{12}\left[ \langle a, c_{d-1}\rangle \langle b, \gamma_n\rangle +\langle b, c_{d-1}\rangle \langle a, \gamma_n\rangle\right] \delta'''(x-y)
\nn\\
&&
\nn\\
&&
\left\{ \langle a, {\bf u}(x)\rangle, \langle b , {\bf u}(y)\rangle\right\}_2=
\nn\\
&&
=\left[\left\langle \left( \frac{a}2 +\mu(a)\right)\cdot b, {\bf u}(x) \right\rangle + \left\langle a\cdot \left( \frac{b}2 +\mu(b)\right) , {\bf u}(y)\right\rangle+\left\langle a\cdot b, c_1(X)\right\rangle\right]\delta'(x-y)
\nn\\
&&
-\frac{\epsilon^2}{12} \left[ \pal_x \left( \langle a, c_{d-1}\rangle \langle b\cdot \gamma_n, {\bf u}(x)\rangle \delta''(x-y)\right) -\pal_y\left( \langle b, c_{d-1}\rangle \langle a\cdot \gamma_n, {\bf u}(y)\rangle \delta''(x-y)\right)\right]
\nn\\
&&
+\frac{\epsilon^2}{12}  \left\langle a\cdot b, \frac{3-d}2 c_d -c_1\cdot c_{d-1}\right\rangle\delta'''(x-y).
\eeqa

For $d<4$ some degree 0 intersection numbers of genus $g>1$ can be nonzero. We will consider just one very simple example of $X={\bf P}^1$ from which the following expressions for intersection numbers of genus 2 tautological classes can be derived.

\begin{prop} The following formula holds true
\beq\label{lambda1}
\sum_{n\geq 1}\frac1{n!}\sum_{k_1, \dots, k_n} t_{k_1}\dots t_{k_n}\int_{\overline{\mathcal M}_{2,n}}\psi_1^{k_1}\dots \psi_n^{k_n}\lambda_1 =\frac1{480} \frac{v_{xxx}}{v_x} -\frac{11}{5760} \frac{v_{xx}^2}{v_x^2}
\eeq
Here
$
v({\bf t})$
is the solution \eqref{kw2} to the dispersionless KdV hierarchy.
\end{prop}

\pf From the expression \eqref{fcp1_2} for the genus 2 GW potential for $X={\bf P}^1$ in the degree 0 limit $q\to 0$ one obtains the expression for the generating function of the genus 2 degree 0 invariants
\beq\label{f20}
{\mathcal F}_2^0 = \frac{7}{5760} u_{xx} +\frac{11 v_{xx}^2}{2880 v_x^2} - \frac{v_{xxx}}{240 v_x}.
\eeq
Here $v=v({\bf t})$, $u=u({\bf t}, {\bf s})$ is the topological solution to the dispersionless Principal Hierarchy
\eqa\label{dprincp1}
&&
\frac{\pal v}{\pal t_k} =\frac{\pal }{\pal x} \frac{v^{k+1}}{(k+1)!}, \quad \frac{\pal u}{\pal t_k}=\frac{\pal }{\pal x} \left(\frac{v^{k}}{k!}u\right)
\nn\\
&&
\\
&&
\frac{\pal v}{\pal s_k} =0, \quad  \frac{\pal u}{\pal s_k}=\frac{\pal v}{\pal t_k}.
\nn
\eeqa
Like in Example \ref{cp1} we redenote $t^1_k\to t_k$, $t^2_k\to s_k$. Observe that $v({\bf t})$ coincides with the solution \eqref{kw2} to the dispersionless hierarchy \eqref{hier0}. The function $u=u({\bf t}, {\bf s})$ is a linear homogeneous function of ${\bf s}$.
In our case the Euler class of the obstruction bundle equals \cite{gp98}
\beq\label{euler2}
e({\mathcal E})=\lambda_2 \boxtimes 1 -\lambda_1\boxtimes c_1({\bf P}^1) .
\eeq
So one has the following expression for the genus 2 intersection numbers on the moduli stacks $\overline M_{2,n}({\bf P}^1, \beta=0)$
\eqa\label{rod2}
&&
\left\langle \tau_{k_1}(\gamma_{\alpha_1})\dots \tau_{k_n}(\gamma_{\alpha_n})\right\rangle_{2, n, \beta=0}=
\\
&&
=\int_{\overline{\mathcal M}_{2,n}} \psi_1^{k_1} \dots \psi_n^{k_n} \lambda_2 \int_{{\bf P}^1}  \gamma_{\alpha_1} \dots \gamma_{\alpha_n}-\int_{\overline{\mathcal M}_{2,n}} \psi_1^{k_1} \dots \psi_n^{k_n} \lambda_1 \int_{{\bf P^1} }c_{1}({\bf P}^1) \gamma_{\alpha_1} \dots \gamma_{\alpha_n}.
\nn
\eeqa
Specializing at $\gamma_1=\dots=\gamma_n=1$ one obtains
$$
\frac{\pal^n{\mathcal F}_2^0}{\pal t_{k_1}\dots \pal t_{k_n}}|_{{\bf t}={\bf s}=0}=\left\langle \tau_{k_1}(1)\dots \tau_{k_n}(1)\right\rangle_{2, n, \beta=0}=-2 \int_{\overline{\mathcal M}_{2,n}} \psi_1^{k_1} \dots \psi_n^{k_n} \lambda_1 .
$$
Since the derivatives of $u$ all vanish at ${\bf s}=0$ the above expression together with \eqref{f20} immediately implies \eqref{euler2}. \epf

\begin{remark} With the help of formula \eqref{f20} one can easily derive also the formula of C.Faber and R.Pandharipande \cite{fp} for the following intersection numbers
\beq\label{lambda2}
\int_{\overline{\mathcal M}_{2,n}}\psi_1^{k_1}\dots \psi_n^{k_n}\lambda_2=\frac7{5760} \frac{(n+1)!}{k_1!\dots k_n!}.
\eeq
Note that this number equals zero unless
$$
k_1+\dots+k_n=n+1.
$$ 

To this end let us choose, say, $\gamma_{\alpha_1}=\omega$ and others $\gamma_{\alpha_2}$, \dots, $\gamma_{\alpha_n}$ equal to 1. The formula \eqref{rod2} together with \eqref{f20} yield
\eqa
&&
\int_{\overline{\mathcal M}_{2,n}} \psi_1^{k_1}\dots \psi_n^{k_n}\lambda_2 =\frac7{5760} \pal_x^2 \frac{\pal^n u}{\pal s_{k_1} \pal t_{k_2} \dots \pal t_{k_n}}|_{{\bf t}=0}=\frac7{5760} \pal_x^2 \frac{\pal^n v}{\pal t_{k_1} \pal t_{k_2} \dots \pal t_{k_n}}|_{{\bf t}=0}=
\nn\\
&&
= \frac7{5760} \left(\pal_x^{n+2} \frac{v^{k_1+\dots+k_n+1}}{(k_1+\dots+k_n+1) k_1!\dots k_n!}\right)_{{\bf t}=0}
=
\frac7{5760} \frac{(n+1)!}{k_1!\dots k_n!}.
\nn
\eeqa
\end{remark}

In conclusion of this section let us compare the brackets \eqref{Biham1} with \eqref{cmp98} considering the second Poisson bracket of the variable
$$
u_1=\int_X {\bf u}.
$$
From \eqref{Biham1} one readily obtains
\beq\label{Biham12}
\{ u_1(x), u_1(y)\}_2=\frac{1-d}2 \left[ u_1(x)+u_1(y)\right]\delta'(x-y)+\frac{\epsilon^2}{12}\left[ \frac{3-d}2 n -\langle c_1, c_{d-1}\rangle\right] \delta'''(x-y).
\eeq
As the expression \eqref{cmp98} does not depend on $q$, one must have
$$
\frac{1}{12}\left[ \frac{3-d}2 n -\langle c_1, c_{d-1}\rangle\right]=\frac{1}2 \tr\left[ \frac14 -\mu^2\right].
$$
We arrive at the following
constraint on a Chern number of a variety with semisimple quantum cohomology.

\begin{prop} \label{prop45} For a smooth projective variety $X$ with $H^{\rm odd}(X)=0$ and with semisimple quantum cohomology the following equation holds true
\beq\label{chern}
\int_X c_1\wedge c_{d-1} =\frac32 \tr \deg^2 -\frac{\chi}2 \, \dim\, (3\dim +1).
%6 \,{\rm tr}\,\mu^2-\frac12n\, d.
\eeq
Here $\chi =\chi(X)$ the Euler characteristic, $\dim=\dim_{\mathbb C}X$.
\end{prop}

\begin{remark} After the first version of this paper was submitted to the archive the author was informed by Hsian-Hua Tseng and Burt Totaro about results of the papers \cite{libgober} and \cite{borisov}. 
In \cite{libgober} it was proven that the Chern number $\int_X c_1\wedge c_{d-1}$ of any $d$-dimensional K\"ahler manifold $X$ can be expressed via its Hodge numbers.
In the paper \cite{borisov} stimulated by \cite{ehx} it was shown that equation \eqref{chern} for a variety $X$ with trivial odd cohomology holds true if an only if the Hodge numbers $h^{p,q}(X)$ all vanish for $p\neq q$. Vanishing of these numbers for smooth projective varieties with semisimple quantum cohomology was proven by A.Bayer and Yu.I.Manin \cite{bayer-manin}. It would be interesting to analyze the possibility of extending the construction of integrable hierarchies of topological type for smooth projective varieties with vanishing non-diagonal Hodge numbers but with non-semisimple quantum cohomology. We plan to do it in subsequent publications.
\end{remark}

%One can easily check validity of eq. \eqref{chern} for various simple examples of Fano varieties with vanishing $H^{\rm odd}(X)$ and semisimple quantum cohomology. For example, for projective space $X={\bf P}^d$ one has $\chi=n=d+1$, 
%$$
%\deg =\diag\left(0, 2, 4,\dots, 2(d-1), 2d\right),
%$$
%so
%$$
%\tr \deg^2=\frac23 d(d+1) (2d+1).
%$$
%Next,
%$$
%c_1(X) = (d+1) \omega, \quad c_{d-1}(X) =\frac{d(d+1)}2\omega^{d-1}
%$$
%where $\omega\in H^2 (X)$ is the K\"ahler form normalized by
%$$
%\int_X\omega^d=1.
%$$
%One arrives at an elementary identity
%$$
%\frac{d(d+1)^2}2 =d(d+1)(2d+1) - \frac12 d(d+1)(3d+1).
%$$
%It is also easy to show that eq. \eqref{chern} is compatible with Cartesian products of varieties, i.e., if it holds true for $X$ and $Y$ then it remains valid for $X\times Y$. Needless to say that the condition of semisimplicity of quantum cohomology is also compatible with Cartesian product (see more details in \cite{kaufm}). A larger class of toric varieties deserves a separate investigation.
%It would also be interesting to apply the topological constraint \eqref{chern} for proving that quantum cohomology of certain Fano varieties is not semisimple. We will analyze further examples in subsequent publications.

\bigskip

{\bf Acknowledgments.} The author is grateful to G.~Borot, H.~Iritani, Y.~Zhang for fruitful discussions and to D.~Orlov for help with checking validity of the constraint \eqref{chern} for certain Fano varieties. This work is
partially supported by the European Research Council Advanced Grant FroM-PDE, by the Russian Federation Government Grant No. 2010-220-01-077 and by PRIN 2010-11 Grant ``Geometric and analytic theory of Hamiltonian systems in finite and infinite dimensions'' of Italian Ministry of Universities and Researches.

\end{document}